\documentclass[%
onecolumn,
amsmath,amssymb,
nofootinbib,
showkeys,
]{revtex4-2}

\usepackage{graphicx}
\usepackage{bm}

\begin{document}

\title{Noether charge formalism for Weyl invariant theories of gravity}

\author{A. Alonso-Serrano}
\email{ana.alonso.serrano@aei.mpg.de}
\affiliation{Max-Planck-Institut f\"ur Gravitationsphysik (Albert-Einstein-Institut), \\Am M\"{u}hlenberg 1, 14476 Potsdam, Germany}

\author{L. J. Garay}
\email{luisj.garay@ucm.es}
\affiliation{Departamento de F\'{i}sica Te\'{o}rica and IPARCOS, Universidad Complutense de Madrid, 28040 Madrid, Spain}

\author{M. Li\v{s}ka,}
\email{liska.mk@seznam.cz}
\affiliation{Institute of Theoretical Physics, Faculty of Mathematics and Physics, Charles University,
	V Hole\v{s}ovi\v{c}k\'{a}ch 2, 180 00 Prague 8, Czech Republic}
\affiliation{Max-Planck-Institut  f\"ur  Gravitationsphysik  (Albert-Einstein-Institut), \\
	Am M\"{u}hlenberg 1, 14476 Potsdam, Germany}

\newcommand{\red}[1]{{\slshape\color{red} #1}}
\newcommand{\blue}[1]{{\slshape\color{blue} #1}}

\begin{abstract}
Gravitational theories invariant under transverse diffeomorphisms and Weyl transformations have the same classical solutions as the corresponding fully diffeomorphism invariant theories. However, they solve some of the problems related to the cosmological constant and in principle allow local energy non-conservation. In the present work, we obtain the Noether charge formalism for these theories. We first derive expressions for the Noether currents and charges corresponding to transverse diffeomorphisms and Weyl transformations, showing that the latter vanish identically. We then use these results to obtain an expression for a perturbation of a Hamiltonian corresponding to evolution along a transverse diffeomorphism generator. From this expression, we derive the first law of black hole mechanics, identifying the total energy, the total angular momentum, the Wald entropy, and the contributions of the cosmological constant perturbations and energy non-conservation. Lastly, we extend our formalism to derive the first law of causal diamonds.
\end{abstract}

\maketitle

\section{Introduction}

The Noether charge formalism for gravitational theories offers a systematic way to calculate conserved quantities without relying on the specifics of the theory~\cite{Wald:1990,Wald:1993,Wald:1994,Iyer:1996,Wald:2000}. In its original form, this method is applicable to any local, diffeomorphism (Diff) invariant theory of gravity. It has been used to derive the first law of black hole mechanics and a prescription for black hole entropy (Wald entropy)~\cite{Wald:1993,Wald:1994,Iyer:1996}, as well as to find the conserved charges corresponding to the Bondi-Metzner-Sachs symmetry group~\cite{Wald:2000}.

We have recently presented an extension of the Noether charge formalism valid for Weyl transverse gravity~\cite{Alonso:2022}. This theory has the same classical solutions as general relativity, but rather than being Diff invariant, its symmetries are transverse diffeomorphisms and Weyl transformations (WTDiff). It emerged from the observation that constructing a consistent theory of self-interacting gravitons can lead either to general relativity or Weyl transverse gravity, depending on the choice of the symmetry group~\cite{Alvarez:2006,Barcelo:2014,Barcelo:2018}. The interest in Weyl transverse gravity comes also from the fact that it solves some of the problems connected with the cosmological constant in general relativity~\cite{Weinberg:1989,Polchinski:2006,Burgess:2013}. In particular, in Weyl transverse gravity the vacuum energy does not gravitate, the cosmological constant appears as an integration constant, and its value is radiatively stable~\cite{Barcelo:2014,Carballo:2015,Barcelo:2018}.

Here, we generalise the framework we developed for Weyl transverse gravity~\cite{Alonso:2022} and obtain the Noether charge formalism applicable to any local, WTDiff-invariant gravitational theory. While the literature concerned with WTDiff-invariant gravity mostly deals with the simplest case, Weyl transverse gravity, there exists a wide class of WTDiff-invariant gravitational theories. In fact, it can be shown that for every Diff-invariant theory there exists a WTDiff-invariant one with the same classical solutions~\cite{Garcia-Moreno}. We review the main features of these theories in section~\ref{WTDiff}. Then, in section~\ref{Noether} we derive the Noether currents and charges corresponding to the WTDiff symmetries and expressions for the symplectic form. In section~\ref{physics}, we apply these general results to obtain a formulation of the first law of black hole mechanics and an expression for Wald entropy for a general local, WTDiff-invariant gravitational theory. In section~\ref{conformal}, we discuss the application of our formalism to causal diamonds, which highlights some differences between WTDiff and Diff-invariant theories of gravity. Lastly, section~\ref{discussion} sums up our findings and discusses possible future developments. Although the present paper is self-contained, it is closely related to the previous work of the authors specialised to Weyl transverse gravity~\cite{Alonso:2022} and further builds on its results.

Unless otherwise specified, we consider an arbitrary spacetime dimension $n$ and metric signature $(-,+,...,+)$. We set $G=c=\hbar=k_{\text{B}}=1$. Definitions of the curvature-related quantities follow~\cite{MTW}.

\section{WTDiff-invariant theories of gravity}
\label{WTDiff}

In this section, we review the properties of the most general local, WTDiff-invariant theories of gravity. We start by discussing the vacuum case and then review how does WTDiff-invariant gravity couple to matter fields.

\subsection{Vacuum WTDiff-invariant gravity}

To construct a WTDiff-invariant theory of gravity we must consider a spacetime with a nondynamical background volume $n$-form, $\boldsymbol{\omega}=\omega\left(x\right)\text{d}x^{0}\wedge\text{d}x^{1}\wedge...\wedge\text{d}x^{n-1}$, where $\omega\left(x\right)$ is a strictly positive function\footnote{In principle, it is possible to introduce dynamics for $\boldsymbol{\omega}$ without spoiling the salient features of WTDiff-invariant gravity~\cite{Carballo:2020}. However, such dynamics plays no role for our current purposes. Hence, we treat $\boldsymbol{\omega}$ as nondynamical.}~\cite{Barcelo:2018}. Using $\boldsymbol{\omega}$, we define an auxiliary metric 
\begin{equation}
\tilde{g}_{\mu\nu}=\left(\sqrt{-\mathfrak{g}}/\omega\right)^{-2/n}g_{\mu\nu},
\end{equation}
where $\mathfrak{g}$ denotes the determinant of the metric. The auxiliary metric $\tilde{g}_{\mu\nu}$ is WTDiff invariant but not Diff invariant. In fact, $\tilde{g}_{\mu\nu}$ is simply the dynamical metric, $g_{\mu\nu}$, restricted to the unimodular gauge given by the condition $\sqrt{-\mathfrak{g}}=\omega$. While one may formulate WTDiff-invariant gravity completely in the unimodular gauge and use $\tilde{g}_{\mu\nu}$ as the dynamical variable, we impose no gauge restrictions and treat $\tilde{g}_{\mu\nu}$ as a mere notational device. Raising and lowering of indices is always performed with the dynamical metric, $g_{\mu\nu}$ (with the obvious exception of $\tilde{g}^{\mu\nu}$ defined as an inverse of $\tilde{g}_{\mu\nu}$). An auxiliary Weyl connection $\tilde{\Gamma}^{\mu}_{\;\:\nu\rho}$ that is Levi-Civita with respect to $\tilde{g}_{\mu\nu}$, i.e., $\tilde{\Gamma}^{\mu}_{\;\:\nu\rho}=\tilde{\Gamma}^{\mu}_{\;\:(\nu\rho)}$ and $\tilde{\nabla}_{\rho}\tilde{g}_{\mu\nu}=0$, reads
\begin{equation}
\tilde{\Gamma}^{\mu}_{\;\:\nu\rho}=\Gamma^{\mu}_{\;\:\nu\rho}-\frac{1}{n}\left(\delta^{\mu}_{\nu}\delta^{\alpha}_{\rho}+\delta^{\mu}_{\rho}\delta^{\alpha}_{\nu}-g_{\nu\rho}g^{\mu\alpha}\right)\partial_{\alpha}\ln\frac{\sqrt{-\mathfrak{g}}}{\omega}.
\end{equation}
The corresponding Riemann tensor equals
\begin{align}
\nonumber \tilde{R}^{\mu}_{\;\:\nu\rho\sigma}=& 2\tilde{\Gamma}^{\mu}_{\;\:\nu[\sigma,\rho]}+2\tilde{\Gamma}^{\mu}_{\;\:\lambda[\rho}\tilde{\Gamma}^{\lambda}_{\;\:\sigma]\nu}\\
\nonumber =&R^{\mu}_{\;\:\nu\rho\sigma}+\frac{2}{n}\left(\delta^{\alpha}_{\nu}\delta^{\mu}_{[\rho}\delta^{\beta}_{\sigma]}-g^{\mu\alpha}g_{\nu[\rho}\delta^{\beta}_{\sigma]}\right)\nabla_{\alpha}\nabla_{\beta}\ln\frac{\sqrt{-\mathfrak{g}}}{\omega} \\
&+\frac{2}{n^2}\left(g^{\alpha\beta}g_{\nu[\rho}\delta^{\mu}_{\sigma]}-3\delta^{\alpha}_{\nu}\delta^{\mu}_{[\rho}\delta^{\beta}_{\sigma]}+3g^{\mu\alpha}g_{\nu[\rho}\delta^{\beta}_{\sigma]}\right)\nabla_{\alpha}\ln\frac{\sqrt{-\mathfrak{g}}}{\omega}\nabla_{\beta}\ln\frac{\sqrt{-\mathfrak{g}}}{\omega}.
\end{align}
These auxiliary quantities allow us to write an action for the most general vacuum, local, WTDiff-invariant theory of gravity,
\begin{equation}
\label{S WTDiff}
S=\int_{V}L\left(\tilde{g}_{\mu\nu}, \tilde{R}^{\mu}_{\;\:\nu\rho\sigma},\tilde{\nabla}_{\alpha_1}\tilde{R}^{\mu}_{\;\:\nu\rho\sigma},..., \tilde{\nabla}_{(\alpha_1}...\tilde{\nabla}_{\alpha_{p})}\tilde{R}^{\mu}_{\;\:\nu\rho\sigma}\right)\omega\text{d}^{n}x,
\end{equation}
where $V$ denotes the spacetime region and $p$ is a natural number. We can limit ourselves to terms with fully symmetrised derivatives, as a general $k$-th Weyl covariant derivative can be rewritten as a fully symmetrised one plus lower derivative terms involving the auxiliary Riemann tensor. By construction, $S$ is invariant with respect to Weyl transformations
\begin{equation}
\delta g_{\mu\nu}=e^{2\sigma}g_{\mu\nu},
\end{equation}
where $\sigma$ denotes an arbitrary spacetime function. The action is also invariant under transverse diffeomorphisms. To make the transversality condition on the diffeomorphism generator $\xi^{\mu}$ Weyl invariant, we must state it with respect to the Weyl covariant derivative,
\begin{align}
\delta g_{\mu\nu}=&2\nabla_{(\nu}\xi_{\mu)}, \\
\tilde{\nabla}_{\mu}\xi^{\mu}=&0\qquad\iff\qquad\nabla_{\mu}\xi^{\mu}=\xi^{\mu}\partial_{\mu}\ln\frac{\sqrt{-\mathfrak{g}}}{\omega}.
\end{align}
Clearly, the WTDiff-invariant actions are not invariant under diffeomorphisms with a nonzero longitudinal component, i.e., those with $\tilde{\nabla}_{\mu}\xi^{\mu}\ne0$.

Varying the action~\eqref{S WTDiff} with respect to $g_{\mu\nu}$ yields the traceless vacuum equations of motion (we will use the symbol $\circ$ for traceless, WTDiff-invariant tensors)
\begin{equation}
16\pi\left(\frac{\sqrt{-\mathfrak{g}}}{\omega}\right)^{\frac{2}{n}}\frac{\delta S}{\delta g_{\mu\nu}}=\mathring{A}^{\mu\nu}=0.
\end{equation}
The factor $\left(\sqrt{-\mathfrak{g}}/\omega\right)^{2/n}$ is added to the left hand side to define the equations of motion as being Weyl invariant. For illustration, in the case of vacuum Weyl transverse gravity, $\mathring{A}^{\mu\nu}$ is just minus the traceless part of the WTDiff-invariant (auxiliary) Ricci tensor. The overall sign of the equations of motion is opposite to the usual one because we vary the action with respect to $g_{\mu\nu}$ rather than $g^{\mu\nu}$.

The invariance of $S$ with respect to transverse diffeomorphisms requires
\begin{align}
\label{Bianchi}
\nonumber 0=&\frac{\delta S}{\delta g_{\mu\nu}}\nabla_{(\nu}\xi_{\mu)}=\frac{1}{16\pi}\int_{V}\left(\frac{\sqrt{-\mathfrak{g}}}{\omega}\right)^{-\frac{2}{n}}\mathring{A}^{\mu\nu}\nabla_{(\nu}\xi_{\mu)}\omega\text{d}^{n}x \\
=&\frac{1}{16\pi}\int_{V}\left(\frac{\sqrt{-\mathfrak{g}}}{\omega}\right)^{-\frac{2}{n}}\xi_{\mu}\tilde{\nabla}_{\nu}\mathring{A}^{\mu\nu}\omega\text{d}^{n}x,
\end{align}
where we impose suitable boundary conditions on the metric and its derivatives to ensure that all the boundary integrals vanish\footnote{We briefly explain the use of the Gauss theorem to convert an integral of a Weyl covariant divergence to a boundary integral. Consider an integral of a Weyl covariant divergence of any Weyl invariant vector $W^{\mu}$. We have
\begin{equation}
\nonumber
\int_{V}\tilde{\nabla}_{\mu}W^{\mu}\omega_{\alpha_1\dots\alpha_n}=\int_{V}\left(\omega\partial_{\mu}W^{\mu}+W^{\mu}\partial_{\mu}\omega\right)\epsilon_{\alpha_1\dots\alpha_n}=\int_{V}\partial_{\mu}\left(\omega W^{\mu}\right)\epsilon_{\alpha_1\dots\alpha_n},
\end{equation}
where $\epsilon_{\alpha_1\dots\alpha_n}$ denotes the $n$-dimensional antisymmetrisation symbol and $\omega_{\alpha_1\dots\alpha_n}=\omega\epsilon_{\alpha_1\dots\alpha_n}$ gives the background volume element. Applying the Gauss theorem yields
\begin{equation}
\nonumber
\int_{V}\partial_{\mu}\left(\omega W^{\mu}\right)\text{d}^{n}x=\int_{\partial V}W^{\mu}n_{\mu}n^{\alpha_1}\omega_{\alpha_1\dots\alpha_n},
\end{equation}
Here, $n^{\mu}$ is a unit normal to $\partial V$, which transforms as $n'^{\mu}=e^{-\sigma}n^{\mu}$ under Weyl transformations. If $n^{\mu}$ corresponds to a coordinate vector, we have
\begin{equation}
\nonumber
\int_{\partial V}W^{\mu}n_{\mu}n^{\alpha_1}\omega_{\alpha_1\dots\alpha_n}=\int_{\partial V}\left(\sqrt{-\mathfrak{g}}/\omega\right)^{-1/n}W^{\mu}n_{\mu}\text{d}^{n-1}x.
\end{equation}}. Since equation~\eqref{Bianchi} holds for any $\xi^{\mu}$ satisfying the transversality condition, it implies
\begin{equation}
\tilde{\nabla}_{\nu}\mathring{A}^{\mu\nu}=\tilde{\nabla}_{\mu}\Phi,
\end{equation}
for some WTDiff-invariant scalar $\Phi$ (the form of $\Phi$ can be explicitly determined, up to a constant, by rewriting the divergence of the traceless equations of motion as a gradient of a scalar). Therefore, we have on shell $\Phi=-\Lambda$, where $\Lambda$ is an arbitrary integration constant. The divergence-free equations of motion then read\footnote{In Weyl transverse gravity, we have $\Phi=-\tilde{R}\left(n-2\right)/2n$, and the divergence-free equations of motion are
\begin{equation}
\nonumber
-\tilde{g}_{\mu\rho}\tilde{g}_{\nu\sigma}\tilde{R}^{\rho\sigma}+\frac{1}{2}\tilde{R}\tilde{g}^{\mu\nu}+\Lambda\tilde{g}^{\mu\nu}=0.
\end{equation}}
\begin{equation}
\label{divergence-free}
\mathring{A}^{\mu\nu}-\left(\Phi+\Lambda\right)\tilde{g}^{\mu\nu}=0.
\end{equation}
In the unimodular gauge, $\sqrt{-\mathfrak{g}}=\omega$, these equations reduce to the equations of motion of some Diff-invariant theory, which has the same classical solutions~\cite{Garcia-Moreno}. We can see that $\Lambda$ plays the role of the cosmological constant. Rather than being a fixed parameter in the Lagrangian, $\Lambda$ appears as an integration constant in the process of solving the equations of motion. Hence, its value can in principle be different for each solution.

\subsection{Coupling to matter fields}
\label{matter}

Upon introducing the vacuum WTDiff-invariant gravity, we discuss its coupling to matter fields. We first introduce the appropriate action for minimally coupled matter fields. Then, we discuss the equations of motion and the question of local energy-momentum conservation. Lastly, we introduce a general local, WTDiff-invariant action which allows non-minimal coupling of matter fields to gravity.

For a matter field minimally coupled to WTDiff-invariant gravity we have the following action
\begin{equation}
\label{S matter}
S_{\psi}=\int_{V}\left(\frac{\sqrt{-\mathfrak{g}}}{\omega}\right)^{2k/n}L_{\psi}\omega\text{d}^{n}x,
\end{equation}
where $L_{\psi}$ is some function of matter variables $\psi$, their Weyl covariant derivatives and $k$ contravariant metric tensors, $g^{\mu\nu}$. The factor $\left(\sqrt{-\mathfrak{g}}/\omega\right)^{2k/n}$ is present to compensate the behaviour of $g^{\mu\nu}$ under Weyl transformations and make the action Weyl invariant (Weyl transformations by definition affect only the metric and leave $\psi$ unchanged). If several matter fields are present, the Lagrangian is simply a sum of several terms of the above described form (including interaction terms).

A variation of the matter action with respect to $\psi$ yields the matter equations of motion
\begin{equation}
\label{matter EoM}
A_{\psi}=0,
\end{equation}
where we do not specify the index structure. Varying both the gravitational and matter actions with respect to $g_{\mu\nu}$ then gives us the gravitational equations of motion
\begin{equation}
\label{EoMs}
\mathring{A}^{\mu\nu}=-8\pi\mathring{T}^{\mu\nu},
\end{equation}
where
\begin{equation}
\mathring{T}^{\mu\nu}=\left(\frac{\sqrt{-\mathfrak{g}}}{\omega}\right)^{2\frac{k+1}{n}}\left(T^{\mu\nu}-\frac{1}{n}Tg^{\mu\nu}\right),
\end{equation}
denotes the traceless WTDiff-invariant part of the standard Hilbert energy-momentum tensor\footnote{One could instead define the energy-momentum tensor with respect to the auxiliary metric~\cite{Carballo:2020}
\begin{equation}
\nonumber
\tilde{T}^{\mu\nu}=2\frac{\partial\left[\left(\sqrt{-\mathfrak{g}}/\omega\right)^{2k/n}L_{\psi}\right]}{\partial\tilde{g}_{\mu\nu}}+\left(\sqrt{-\mathfrak{g}}/\omega\right)^{2k/n}L_{\psi}\tilde{g}_{\mu\nu},
\end{equation}
making it Weyl invariant. However, the standard definition has the advantage of involving variations with respect to the full dynamical metric.}
\begin{equation}
T^{\mu\nu}=2\frac{\partial L_{\psi}}{\partial g_{\mu\nu}}+L_{\psi}g^{\mu\nu},
\end{equation}
and $T$ being its trace. For Weyl transformations of $T^{\mu\nu}$ we have
\begin{equation}
T'^{\mu\nu}=e^{-2\left(k+1\right)\sigma}T^{\mu\nu},
\end{equation}
and the gravitational equations of motion~\eqref{EoMs} are thus Weyl invariant.

The invariance of the matter action with respect to transverse diffeomorphisms implies~\cite{Alvarez:2013a}
\begin{equation}
\label{non-conservation}
8\pi\tilde{\nabla}_{\nu}\left[\left(\sqrt{-g}/\omega\right)^{2k/n}T_{\mu}^{\;\:\nu}\right]=\tilde{\nabla}_{\mu}\mathcal{J},
\end{equation}
for some scalar function $\mathcal{J}$ defined up to a constant through this equation. $\mathcal{J}\ne0$ corresponds to breaking of the local energy-momentum conservation. While the matter equations of motion~\eqref{matter EoM} suffice to ensure $\mathcal{J}=0$ for many of the often considered matter fields, the freedom to introduce $\mathcal{J}\ne0$ has been used, e.g. as a possible way to account for cosmic acceleration~\cite{Josset:2017,Perez:2018}. Therefore, in the present work, we assume the most general situation, $\mathcal{J}\ne0$, so that the divergence-free equations of motion then include a term proportional to $\mathcal{J}$
\begin{equation}
\mathring{A}^{\mu\nu}-\left(\Phi+\mathcal{J}+\Lambda\right)\tilde{g}^{\mu\nu}=-8\pi\left(\frac{\sqrt{-\mathfrak{g}}}{\omega}\right)^{2\frac{k+1}{n}}T^{\mu\nu}.
\end{equation}

In the following, we will consider a more general class of WTDiff-invariant theories. Their Lagrangian can be any scalar built from the auxiliary metric, $\tilde{g}_{\mu\nu}$, the corresponding Riemann tensor, $\tilde{R}^{\mu}_{\;\:\nu\rho\sigma}$ and at most its  $p$-th Weyl covariant derivatives and some collection of matter fields, collectively denoted by $\psi$, and at most their $q$-th  Weyl covariant derivatives ($p$ and $q$ are arbitrary natural numbers). We also allow non-minimal coupling of the matter fields to gravity. The corresponding action reads
\begin{equation}
\label{action}
S=\int_{\Omega}L\left(\tilde{g}_{\mu\nu}, \tilde{R}^{\mu}_{\;\:\nu\rho\sigma},\tilde{\nabla}_{\alpha_1}\tilde{R}^{\mu}_{\;\:\nu\rho\sigma},..., \tilde{\nabla}_{(\alpha_1}...\tilde{\nabla}_{\alpha_{p})}\tilde{R}^{\mu}_{\;\:\nu\rho\sigma}, \psi,\tilde{\nabla}_{\alpha_1}\psi,...,\tilde{\nabla}_{(\alpha_1}...\tilde{\nabla}_{\alpha_{q})}\psi\right)\omega\text{d}^{n}x.
\end{equation}
In this case, we define the energy-momentum tensor as before and include all the terms coupling matter variables to the Riemann tensor and its derivatives on the left hand side of the equations of motion. Thus, $\mathring{A}^{\mu\nu}$ can in general depend on $\psi$ and its derivatives, whereas $T_{\mu\nu}$ depends only on the metric and not on its derivatives.

One can show that for every local, WTDiff-invariant action which implies local energy-momentum conservation there exists a corresponding local, Diff-invariant action which leads to the same classical gravitational dynamics (except for the different behaviour of $\Lambda$) and vice versa~\cite{Garcia-Moreno}. In other words, we have pairs of Diff and WTDiff-invariant theories of gravity that are mutually equivalent in the same sense as general relativity and Weyl transverse gravity are.

\section{Noether charge formalism}
\label{Noether}

In this section, we develop the Noether charge formalism for any local, WTDiff-invariant theory of gravity with matter sources (not necessarily minimally coupled). We first give a brief overview of the Noether charge formalism in a general context. Specialising to WTDiff-invariant gravity, we derive the symplectic potential and current for general variations of the metric and the matter fields. Then, we consider variations corresponding to local symmetries, i.e., Weyl transformations and transverse diffeomorphisms, and obtain expressions for the Noether currents and charges. Lastly, we express a perturbation of a Hamiltonian corresponding to evolution along a transverse diffeomorphism generator in terms of the Noether charge. This expression will be crucial for deriving the first law of black hole mechanics in the next section.

\subsection{General formalism}
\label{general}

Before going to WTDiff-invariant gravity, we introduce the Noether charge formalism in a general setting. Consider a manifold equipped with a volume form, $\boldsymbol{\varepsilon}$, and a (nonunique) covariant derivative operator satisfying $\nabla_{\mu}\boldsymbol{\varepsilon}=0$. On this manifold define a Lagrangian $L\left[\phi,\gamma\right]$ as a local function of dynamical variables, $\phi$, and nondynamical variables, $\gamma$, and their finitely many covariant derivatives. Under an arbitrary variation of the dynamical variables, $\delta_1 \phi$, the variation of $L$ obeys
\begin{equation}
\label{dL general}
\delta_1 L\left[\phi\right]=A_{\phi}\left[\phi\right]\delta_1 \phi+\nabla_{\mu}\theta^{\mu}\left[\delta_1\right].
\end{equation}
Since, due to the Gauss theorem, $\nabla_{\mu}\theta^{\mu}$ contributes only a boundary integral to the variation of the action, the equations of motion read $A_{\phi}=0$. The vector $\theta^{\mu}\left[\delta\right]$ is known as the symplectic potential~\cite{Wald:1990}.

Next, we perform a second independent arbitrary variation, $\delta_2\phi$. The commutator of the variations acting on the Lagrangian yields
\begin{equation}
\left(\delta_1\delta_2-\delta_2\delta_1\right)L\left[\phi,\gamma\right]=\delta_1\left(A_{\phi}\left[\phi\right]\right)\delta_2 \phi-\delta_2\left(A_{\phi}\left[\phi\right]\right)\delta_1 \phi+\nabla_{\mu}\Omega^{\mu}\left[\delta_{1},\delta_{2}\right],
\end{equation}
with
\begin{equation}
\label{omega}
\Omega^{\mu}\left[\delta_{1},\delta_{2}\right]=\delta_{1}\theta^{\mu}\left[\delta_{2}\right]-\delta_{2}\theta^{\mu}\left[\delta_{1}\right],
\end{equation}
being the symplectic current~\cite{Wald:1990}. If we integrate $\Omega^{\mu}\left[\delta_{1},\delta_{2}\right]$ over an initial data surface, $\mathcal{C}$, we obtain the symplectic form\footnote{The form defined in this way is in general degenerate~\cite{Wald:1990}. To construct a true symplectic form, one must restrict it from the space of field configurations to the phase space. However, this subtlety plays no role in the following.}
\begin{equation}
\label{form}
\Omega\left[\delta_{1},\delta_{2}\right]=\int_{\mathcal{C}}\Omega^{\mu}\left[\delta_{1},\delta_{2}\right]\text{d}\mathcal{C}_{\mu}.
\end{equation}
In the special case of a variation generated by a vector field $\xi^{\mu}$, i.e., $\delta_1\phi=\pounds_{\xi}\phi$, the Hamilton equations of motion imply for an arbitrary variation, $\delta_{2}\phi=\delta\phi$, applied to the Hamiltonian $H_{\xi}$ corresponding to the evolution along $\xi^{\mu}$ (if it exists),
\begin{equation}
\label{hamiltonian general}
\delta H_{\xi}=\Omega\left[\pounds_{\xi},\delta\right].
\end{equation}

If the variation we consider corresponds to a local symmetry transformation of the dynamical fields, $\hat{\delta}\phi$, it leads to a variation of the Lagrangian that is a total divergence, \mbox{$\hat{\delta}L=\nabla_{\mu}\alpha^{\mu}$} for some vector $\alpha^{\mu}$. The Noether current corresponding to a gauge transformation then obeys~\cite{Wald:1990}
\begin{equation}
\label{j}
j^{\mu}\big[\hat{\delta}\big]=\theta^{\mu}\big[\hat{\delta}\big]-\alpha^{\mu}\big[\hat{\delta}\big].
\end{equation}
It is easy to see that the covariant divergence of $j^{\mu}$ vanishes on-shell, i.e., for $A_{\phi}=0$,
\begin{equation}
\nabla_{\mu}j^{\mu}=\nabla_{\mu}\theta^{\mu}-\nabla_{\mu}\alpha^{\mu}=\nabla_{\mu}\theta^{\mu}-\hat{\delta}L=-A_{\phi}\hat{\delta}\phi.
\end{equation}
Lastly, an integral of $j^{\mu}$ over an initial data surface yields a conserved Noether charge~\cite{Wald:1990}
\begin{equation}
Q=\int_{\mathcal{C}}j^{\mu}\text{d}\mathcal{C}_{\mu}.
\end{equation}

\subsection{WTDiff symplectic potential}
\label{potential}

We now apply the general formalism to the case of local, WTDiff-invariant gravitational theories, starting with obtaining the symplectic potential. For an arbitrary variation of the Lagrangian~\eqref{action}, some lengthy but fairly straightforward manipulations lead to (we discuss the details in Appendix~\ref{technical})
\begin{align}
\nonumber \delta L=&\frac{1}{16\pi}\left(\frac{\sqrt{-\mathfrak{g}}}{\omega}\right)^{-\frac{2}{n}}\left(\mathring{A}^{\mu\nu}+8\pi\mathring{T}^{\mu\nu}\right)\delta g_{\mu\nu}+A_{\psi}\delta\psi+\tilde{\nabla}_{\alpha_1}\Bigg[2\left(\frac{\sqrt{-\mathfrak{g}}}{\omega}\right)^{\frac{2}{n}}E^{\mu\nu\rho\alpha_1}\tilde{\nabla}_{\mu}\delta\tilde{g}_{\nu\rho} \\
&+K^{\alpha_1\mu\nu}\delta\tilde{g}_{\mu\nu}+\sum_{i=2}^{p}M^{\alpha_1\alpha_2...\alpha_i\;\:\nu\rho\sigma}_{\quad\quad\quad\;\mu}\delta\hat{\nabla}_{(\alpha_2}...\tilde{\nabla}_{\alpha_i)}\tilde{R}^{\mu}_{\;\:\nu\rho\sigma}+\sum_{i=2}^{q}N^{\alpha_1\alpha_2...\alpha_i}\delta\tilde{\nabla}_{(\alpha_2}...\hat{\nabla}_{\alpha_i)}\psi\Bigg]. \label{dL}
\end{align}
The term contracted with $\delta g_{\mu\nu}$ corresponds to the gravitational equations of motion~\eqref{EoMs} and the one contracted with $\psi$ to the matter equations of motion~\eqref{matter EoM}. The tensors $E_{\mu}^{\;\:\nu\rho\sigma}$,  $K^{\alpha_1\mu\nu}$, $M^{\alpha_1\alpha_2...\alpha_i\;\:\nu\rho\sigma}_{\quad\quad\quad\;\mu}$, and $N^{\alpha_1\alpha_2...\alpha_i}$ are assumed to have the same symmetries as the expressions they are contracted with. In particular, $E_{\mu}^{\;\:\nu\rho\sigma}$ possesses the symmetries of the Riemann tensor and reads
\begin{equation}
\label{entropy density}
E_{\mu}^{\;\:\nu\rho\sigma}=\sum_{i=0}^{p}\left(-1\right)^{i}\tilde{\nabla}_{\alpha_1}...\tilde{\nabla}_{\alpha_i}\left(\frac{\partial L}{\partial\tilde{\nabla}_{(\alpha_1}...\tilde{\nabla}_{\alpha_i)}\tilde{R}^{\mu}_{\;\:\nu\rho\sigma}}\right).
\end{equation}
The precise forms of $K^{\alpha_1\mu\nu}$, $M^{\alpha_1\alpha_2...\alpha_i\;\:\nu\rho\sigma}_{\quad\quad\quad\;\mu}$ and $N^{\alpha_1\alpha_2...\alpha_i}$ are unimportant for our purposes. It only matters that that they are tensors constructed from $\tilde{g}_{\mu\nu}$, $\tilde{\nabla}_{\mu}$, $\tilde{R}^{\mu}_{\;\:\nu\rho\sigma}$, and $\psi$, and the terms they are contracted with contain no derivatives of variations.

By comparing equations~\eqref{dL general} and~\eqref{dL}, we can easily identify the symplectic potential for WTDiff-invariant theories
\begin{align}
\nonumber \theta^{\mu}\left[\delta\right]=&2\left(\frac{\sqrt{-\mathfrak{g}}}{\omega}\right)^{\frac{2}{n}}E^{\sigma\nu\rho\mu}\tilde{\nabla}_{\sigma}\delta\tilde{g}_{\nu\rho}+K^{\mu\nu\rho}\delta\tilde{g}_{\nu\rho}+\sum_{i=2}^{p}M^{\mu\alpha_2...\alpha_i\;\:\nu\rho\sigma}_{\quad\quad\;\;\;\;\lambda}\delta\tilde{\nabla}_{(\alpha_2}...\tilde{\nabla}_{\alpha_i)}\tilde{R}^{\lambda}_{\;\:\nu\rho\sigma} \\
&+\sum_{i=2}^{q}N^{\mu\alpha_2...\alpha_i}\delta\tilde{\nabla}_{(\alpha_2}...\tilde{\nabla}_{\alpha_i)}\psi. \label{theta}
\end{align}
The symplectic potential is by construction WTDiff invariant. In the unimodular gauge and for metric variations that do not change the determinant, $\delta\mathfrak{g}=0$, equation~\eqref{theta} reduces to the symplectic potential for local, Diff-invariant theories of gravity~\cite{Wald:1994}.

From the symplectic potential we can directly obtain the symplectic current  using the general definition~\eqref{omega}, and, upon integration, the symplectic form, $\Omega\left[\delta_{1},\delta_{2}\right]$. If we evaluate it for a transformation generated by a vector field $\xi^{\mu}$, $\delta_1 g_{\mu\nu}=\delta_{\xi}g_{\mu\nu}$\footnote{In general, $\delta_{\xi}$ applied to a WTDiff-invariant expression does not correspond to a Lie derivative. This is because, on the one hand, the background volume form $\boldsymbol{\omega}$ is nondynamical and, therefore, $\delta_{\xi}\boldsymbol{\omega}=0$. On the other hand, we have $\pounds_{\xi}\boldsymbol{\omega}=\boldsymbol{\omega}\tilde{\nabla}_{\mu}\xi^{\mu}\ne0$. Nevertheless, for transverse diffeomorphisms, on which we will focus in the rest of this section, $\tilde{\nabla}_{\mu}\xi^{\mu}=0$ and it holds $\delta_{\xi}=\pounds_{\xi}$. We return to the general case $\tilde{\nabla}_{\mu}\xi^{\mu}\ne0$ in section~\ref{conformal}.}, and a general metric variation, $\delta_{2}g_{\mu\nu}=\delta g_{\mu\nu}$, we obtain the variation of the Hamiltonian corresponding to the the evolution along $\xi^{\mu}$ (see equation~\eqref{hamiltonian general} and the preceding discussion)
\begin{equation}
\delta H_{\xi}=\Omega\left[\delta_{\xi},\delta\right].
\end{equation}
Finding $\delta H_{\xi}$ represents one of our main goals. However, while the straightforward evaluation is in principle possible, it does not yield an easily interpretable result. For that, one would have to consider a Lagrangian of some specific theory (see section 6 of~\cite{Alonso:2022} for an example of such a direct approach). Instead, in the following, we consider an alternative route, that can be applied to Hamiltonians corresponding to generators of the local symmetries of the theory, i.e., the WTDiff group~\cite{Wald:1993,Wald:1994}.

\subsection{WTDiff Noether current}
\label{current}

Up to this point, we were concerned with arbitrary variations. We now specialise to the variations that do not affect the physical content of the theory, i.e., those corresponding to local symmetry transformations. For these, we obtain the Noether currents and charges and then use them to express a variation of the corresponding Hamiltonian. For WTDiff-invariant theories the local symmetry transformations are infinitesimal transverse diffeomorphisms and Weyl transformations.

We first evaluate the Noether current for local infinitesimal Weyl transformations, $\delta_{\text{W}}g_{\mu\nu}=2\sigma g_{\mu\nu}$ and $\delta_{\text{W}}\psi=0$. The variation of the general local, WTDiff-invariant Lagrangian~\eqref{action} vanishes by construction, $\delta_{\text{W}}L=0$. Therefore, the vector $\alpha^{\mu}\left[\delta_{\text{W}}\right]$ present in the general definition of the Noether current~\eqref{j} equals zero. Moreover, it is easy to check that equation~\eqref{theta} for the symplectic potential yields $\theta^{\mu}\left[\delta_{\text{W}}\right]=0$. In total, we obtain $j^{\mu}\left[\delta_{\text{W}}\right]=0$ in any local, WTDiff-invariant theory of gravity  with arbitrary (even non-minimally coupled) matter fields present. In this way, we generalise the previous proofs of the vanishing Noether current corresponding to Weyl symmetry for Weyl transverse gravity~\cite{Oda:2017,Jackiw:2014} and a recent proof for arbitrary WTDiff-invariant theories of gravity in four spacetime dimensions~\cite{Oda:2022}. It has been argued that the vanishing of the Noether current plays an important role in the radiative stability of the cosmological constant and the absence of a conformal anomaly corresponding to local Weyl transformations in WTG~\cite{Oda:2017}. Hence, our result supports the view that these properties should carry over to arbitrary local, WTDiff-invariant theories of gravity.

Next, we derive the Noether current corresponding to a transverse diffeomorphism generated by a vector field $\xi^{\mu}$ (for which we have $\delta_{\xi}=\pounds_{\xi}$), $\delta_{\xi}g_{\mu\nu}=2\nabla_{(\mu}\xi_{\nu)}$, $\delta_{\xi}\psi=\pounds_{\xi}\psi$, $\tilde{\nabla}_{\mu}\xi^{\mu}=0$. For the variation of the Lagrangian, we have
\begin{equation}
\delta_{\xi}L=\xi^{\mu}\tilde{\nabla}_{\mu}L,
\end{equation}
where we used $\tilde{\nabla}_{\mu}\xi^{\mu}=0$. Then, we have $\alpha^{\mu}\left[\pounds_{\xi}\right]=L\xi^{\mu}$ and the Noether current \mbox{$j^{\mu}_{\xi}=j^{\mu}\left[\pounds_{\xi}\right]$} reads
\begin{equation}
\label{j1}
j^{\mu}_{\xi}=\theta^{\mu}\left[\pounds_{\xi}\right]-L\xi^{\mu}.
\end{equation}
We can obtain a convenient explicit expression for $j^{\mu}_{\xi}$ by studying the divergence of equation~\eqref{j1}. It equals
\begin{align}
\nonumber \tilde{\nabla}_{\mu}j^{\mu}_{\xi}=&\tilde{\nabla}_{\mu}\theta^{\mu}\left[\pounds_{\xi}\right]-\xi^{\mu}\tilde{\nabla}_{\mu}L \\
\nonumber =&-A_{\psi}\pounds_{\xi}\psi-\frac{1}{16\pi}\bigg[\left(\sqrt{-\mathfrak{g}}/\omega\right)^{-2/n}\left(\mathring{A}^{\mu\nu}+8\pi\mathring{T}^{\mu\nu}\right)\bigg]\pounds_{\xi}g_{\mu\nu}+\pounds_{\xi}L-\xi^{\mu}\tilde{\nabla}_{\mu}L \\
=&-\tilde{\nabla}_{\mu}\left(\psi\cdot A_{\psi}\cdot\xi\right)^{\mu}-\frac{1}{16\pi}\bigg[\left(\sqrt{-\mathfrak{g}}/\omega\right)^{-2/n}\left(\mathring{A}^{\mu\nu}+8\pi\mathring{T}^{\mu\nu}\right)\bigg]\pounds_{\xi}g_{\mu\nu}, \label{nabla j}
\end{align}
where we used the general definition of the symplectic current~\eqref{dL general} to get the second equality. The expression $\left(\psi\cdot A_{\psi}\cdot\xi\right)^{\mu}$ stands for $\psi A_{\psi}\xi^{\mu}$ for scalar fields and $2\psi_{\nu}A_{\psi}^{(\mu}\xi^{\nu)}$ for vector fields (in the case of more general tensorial fields, a separate treatment is required).

As required for a Noether current, $\tilde{\nabla}_{\mu}j^{\mu}_{\xi}$ vanishes on shell. Off shell, we can use $\pounds_{\xi}g_{\mu\nu}=2\nabla_{(\mu}\xi_{\nu)}$ to rewrite the last term in equation~\eqref{nabla j} in the following way
\begin{align}
&-\tilde{\nabla}_{\nu}\left[\xi_{\mu}\frac{1}{8\pi}\left(\sqrt{-\mathfrak{g}}/\omega\right)^{-2/n}\left(\mathring{A}^{\mu\nu}+8\pi\mathring{T}^{\mu\nu}\right)\right]+\xi^{\mu}\tilde{\nabla}_{\nu}\left[\frac{1}{8\pi}\left(\sqrt{-\mathfrak{g}}/\omega\right)^{-2/n}\left(\mathring{A}^{\mu\nu}+8\pi\mathring{T}^{\mu\nu}\right)\right].
\end{align}
By virtue of equations~\eqref{Bianchi} and~\eqref{non-conservation} (which both hold off shell) the second term also corresponds to a total divergence, $\tilde{\nabla}_{\mu}\left[\left(1/8\pi\right)\left(\Phi+\mathcal{J}\right)\xi^{\mu}\right]$ (recall that $\tilde{\nabla}_{\mu}\xi^{\mu}=0$). In total, we have shown that $\tilde{\nabla}_{\mu}j^{\mu}_{\xi}$ can be written as a Weyl covariant divergence of some expression. It follows that $j^{\mu}_{\xi}$ must be equal to this expression~\cite{Wald:1990b}, up to a divergence of some WTDiff-invariant antisymmetric tensor, $Q^{\nu\mu}_{\xi}$, and a term of the form $\lambda\xi^{\mu}$, with $\lambda$ being an arbitrary constant (it is easy to check that $\tilde{\nabla}_{\mu}\tilde{\nabla}_{\nu}Q^{\nu\mu}_{\xi}=0$ and, therefore, it does not contribute to $\tilde{\nabla}_{\mu}j^{\mu}_{\xi}$). The term $\lambda\xi^{\mu}$ has been analysed by the authors in the case of Weyl transverse gravity and it has been found that we can set $\lambda=0$ without any loss of generality~\cite{Alonso:2022}. Since the argument works without any changes even for general WTDiff-invariant theories, we do not repeat it here (see subsections 3.3 and 4.3 of~\cite{Alonso:2022} for details). Hence, we find that the Noether current corresponding to a transverse diffeomorphism generated by a vector field $\xi^{\mu}$ can be written as
\begin{equation} j^{\mu}_{\xi}=-\left[\frac{1}{8\pi}\left(\frac{\sqrt{-\mathfrak{g}}}{\omega}\right)^{-\frac{2}{n}}\left(\mathring{A}_{\nu}^{\;\:\mu}+8\pi\mathring{T}_{\nu}^{\;\:\mu}\right)-\frac{1}{8\pi}\left(\Phi+\mathcal{J}\right)\delta_{\nu}^{\mu}\right]\xi^{\nu}-\left(A_{\psi}\cdot\psi\cdot\xi\right)^{\mu}+\tilde{\nabla}_{\nu}Q^{\nu\mu}_{\xi}. \label{j2}
\end{equation}
The antisymmetric tensor $Q^{\nu\mu}_{\xi}$ corresponds to the Noether charge of the transverse diffeomorphism generated by the vector field $\xi^{\mu}$.

For future use, we note that, on shell, the Noether current reduces to
\begin{equation}
\label{on shell current}
j^{\mu}_{\xi}=-\frac{1}{8\pi}\Lambda\xi^{\mu}+\tilde{\nabla}_{\nu}Q^{\nu\mu}_{\xi},
\end{equation}
where $\Lambda$ is an integration constant corresponding to the cosmological constant.

The structure of $Q^{\nu\mu}_{\xi}$ can be partially read off from equation~\eqref{theta} defining the symplectic potential~\cite{Wald:1994,Iyer:1996}. For a transverse diffeomorphism, variations correspond to Lie derivatives with respect to the diffeomorphism generator, $\xi^{\mu}$. From the well-known properties of the Lie derivative it follows that only the term $2E^{\sigma\nu\rho\mu}\tilde{\nabla}_{\sigma}\left(\pounds_{\xi}\tilde{g}_{\nu\rho}\right)$ involves second derivatives of $\xi^{\mu}$ as it is the only term containing a derivative of a variation. Only this term then leads to a contribution to $Q^{\nu\mu}_{\xi}$ proportional to $\tilde{\nabla}^{\nu}\xi^{\mu}$. Hence, it holds
\begin{equation}
\label{charge}
Q^{\nu\mu}_{\xi}=2E^{\nu\mu\rho}_{\quad\;\:\sigma}\tilde{\nabla}_{\rho}\xi^{\sigma}+W_{\rho}^{\;\:\nu\mu}\xi^{\rho},
\end{equation}
where $W_{\rho}^{\;\:\nu\mu}=W_{\rho}^{\;\:[\nu\mu]}$ is some WTDiff invariant tensor independent of $\xi^{\mu}$. An explicit expression for $W_{\rho}^{\;\:\nu\mu}$ depends on the specifics of the Lagrangian.

To conclude, we note that the expressions for the symplectic potentials and currents, as well as the Noether currents and charges contain ambiguities. These are of the same form as for the Diff-invariant theories~\cite{Wald:1994}. Therefore, we do not discuss them in detail and simply point out that they are irrelevant for the physical situations we discuss in the next section. In the following, we will treat all the expressions as effectively unambiguous.

\subsection{Hamiltonian}

We now use expressions~\eqref{j2} and~\eqref{dj1} for $j^{\mu}_{\xi}$ we obtained in the previous subsection to find a variation of the Hamiltonian $H_{\xi}$ corresponding to the evolution along a transverse-diffeomorphism generator $\xi^{\mu}$. Consider an arbitrary variation, $\delta g_{\mu\nu}$, $\delta\psi$. There are two ways to evaluate the corresponding change of $j_{\xi}^{\mu}$ which, upon identification, will allow us to obtain $\delta H_{\xi}$. First, varying equation~\eqref{j2}, which expresses $j^{\mu}_{\xi}$ in terms of the equations of motion and the Noether charge, yields
\begin{equation}
\delta j^{\mu}_{\xi}=-\frac{1}{8\pi}\xi^{\nu}\delta\left[\left(\frac{\sqrt{-\mathfrak{g}}}{\omega}\right)^{-\frac{2}{n}}\mathring{A}_{\nu}^{\;\:\mu}-\left(\Phi+\mathcal{J}\right)\delta_{\nu}^{\mu}+8\pi\left(\frac{\sqrt{-\mathfrak{g}}}{\omega}\right)^{-\frac{2k}{n}}T_{\nu}^{\;\:\mu}\right]-\delta\left(A_{\psi}\cdot\psi\cdot\xi\right)^{\mu}+\tilde{\nabla}_{\nu}\delta Q^{\nu\mu}_{\xi}. \label{dj1}
\end{equation}
Second, we may vary directly the definition of $j^{\mu}_{\xi}$~\eqref{j1}, and express the variation of $L$ in terms of the equations of motion and the symplectic potential
\begin{equation}
\delta j^{\mu}_{\xi}=\delta\theta^{\mu}\left[\pounds_{\xi}\right]-\xi^{\mu}\tilde{\nabla}_{\nu}\theta^{\nu}\left[\delta\right]-\frac{1}{16\pi}\xi^{\mu}\bigg[\left(\sqrt{-\mathfrak{g}}/\omega\right)^{-2/n}\left(\mathring{A}^{\mu\nu}+8\pi\mathring{T}^{\mu\nu}\right)\bigg]\delta g_{\mu\nu}-\xi^{\mu}A_{\psi}\delta\psi. \label{dj2}
\end{equation}
In order to find $\delta H_{\xi}$, we need to relate $\delta j^{\mu}_{\xi}$ with the symplectic current~\eqref{omega} corresponding to the transverse diffeomorphism, $\pounds_{\xi}$, and the arbitrary variation of the metric, $\delta$,
\begin{equation}
\Omega^{\mu}\left[\pounds_{\xi},\delta\right]=\delta\theta^{\mu}\left[\pounds_{\xi}\right]-\pounds_{\xi}\theta^{\mu}\left[\delta\right].
\end{equation}
The first term appears in equation~\eqref{dj2}. We can add and subtract the second term, $\pounds_{\xi}\theta^{\mu}\left[\delta\right]$, to equation~\eqref{dj2}, using that it holds
\begin{equation}
\label{lie theta}
\pounds_{\xi}\theta^{\mu}\left[\delta\right]=\xi^{\nu}\tilde{\nabla}_{\nu}\theta^{\mu}\left[\delta\right]-\theta^{\nu}\left[\delta\right]\tilde{\nabla}_{\nu}\xi^{\mu}.
\end{equation}
Then, we obtain
\begin{align}
\label{dj3}
\delta j^{\mu}_{\xi}=&\Omega^{\mu}\left[\pounds_{\xi},\delta\right]+2\tilde{\nabla}_{\nu}\left(\xi^{[\nu}\theta^{\mu]}\left[\delta\right]\right)-\frac{1}{16\pi}\xi^{\mu}\left(\sqrt{-\mathfrak{g}}/\omega\right)^{-2/n}\left(\mathring{A}^{\mu\nu}+8\pi\mathring{T}^{\mu\nu}\right)\delta g_{\mu\nu}-\xi^{\mu}A_{\psi}\delta\psi.
\end{align}

The other way to calculate $\delta j^{\mu}_{\xi}$ is equation~\eqref{dj1}. If we now equate expressions~\eqref{dj1} and~\eqref{dj3} for $\delta j^{\mu}_{\xi}$, we obtain an expression for the symplectic current and hence for $\delta H_{\xi}$ (if it exists)
\begin{align}
\nonumber \Omega^{\mu}\left[\pounds_{\xi},\delta\right]=&\tilde{\nabla}_{\nu}\left(Q^{\nu\mu}_{\xi}-2\xi^{[\nu}\theta^{\mu]}\left[\delta \right]\right)+\xi^{\mu}A_{\psi}\delta\psi-\delta\left(A_{\psi}\cdot\psi\cdot\xi\right)^{\mu} \\
\nonumber & +\frac{1}{16\pi}\xi^{\mu}\left(\sqrt{-\mathfrak{g}}/\omega\right)^{-2/n}\left(\mathring{A}^{\mu\nu}+8\pi\mathring{T}^{\mu\nu}\right)\delta g_{\mu\nu} \\
&-\xi^{\nu}\delta\bigg[\frac{1}{8\pi}\left(\sqrt{-\mathfrak{g}}/\omega\right)^{-\frac{2}{n}}\mathring{A}_{\nu}^{\mu}-\frac{1}{8\pi}\left(\Phi+\mathcal{J}\right)\delta_{\nu}^{\mu}+\left(\sqrt{-\mathfrak{g}}/\omega\right)^{2\frac{k}{n}}T_{\nu}^{\;\:\mu}\bigg]. \label{dj}
\end{align}
Let us stress that equation~\eqref{dj} holds even off shell.

We now specialise to the physically interesting situation of two solutions of the equations of motion related by a small perturbation. Then, both the background fields $g_{\mu\nu}$, $\psi$ and the perturbations $\delta g_{\mu\nu}$, $\delta\psi$ obey the equations of motion. In particular, we have
\begin{equation}
\delta\left[\frac{1}{8\pi}\left(\frac{\sqrt{-\mathfrak{g}}}{\omega}\right)^{-\frac{2}{n}}\mathring{A}_{\nu}^{\;\:\mu}-\frac{1}{8\pi}\left(\Phi+\mathcal{J}\right)\delta_{\nu}^{\mu}+\left(\frac{\sqrt{-\mathfrak{g}}}{\omega}\right)^{2\frac{k}{n}}T_{\nu}^{\;\:\mu}\right]=\frac{1}{8\pi}\delta\Lambda\delta_{\nu}^{\mu}
\end{equation}
where $\delta\Lambda$ is an arbitrary integration constant. The cosmological constant is then perturbed by $\delta\Lambda$ to a new value, $\Lambda+\delta\Lambda$. In this case, the symplectic current considerably simplifies
\begin{equation}
\Omega^{\mu}\left[\pounds_{\xi},\delta\right]=\tilde{\nabla}_{\nu}\left(\delta Q^{\nu\mu}_{\xi}-2\xi^{[\nu}\theta^{\mu]}\left[\delta\right]\right)-\frac{1}{8\pi}\xi^{\mu}\delta\Lambda.
\end{equation}

Now assume that the unperturbed spacetime possesses a Cauchy surface, $\mathcal{C}$, and introduce a Weyl invariant volume element on it, $\text{d}\mathcal{C}_{\mu}=\left(\sqrt{-\mathfrak{g}}/\omega\right)^{-1/n}n_{\mu}\omega\text{d}^{n-1}x$. Here,  $n_{\mu}$ is the unit normal to $\mathcal{C}$ and $\text{d}^{n-1}x$ the coordinate volume element on $\mathcal{C}$. The Weyl invariant volume element $\text{d}\mathcal{C}_{\mu}$ reduces to the physical one (measured with respect to the dynamical metric) only in the unimodular gauge, $\sqrt{-\mathfrak{g}}=\omega$. The symplectic form $\Omega\left[\pounds_{\xi},\delta\right]$ on $\mathcal{C}$ is then given by an integral of $\Omega^{\mu}\left[\pounds_{\xi},\delta\right]$ over $\mathcal{C}$ (see the general definition~\eqref{form})
\begin{equation}
\label{dH1}
\Omega\left[\pounds_{\xi},\delta\right]=\int_{\mathcal{C}}\Omega^{\mu}\left[\pounds_{\xi},\delta\right]\text{d}\mathcal{C}_{\mu}=\int_{\partial\mathcal{C}}\left(\delta Q^{\nu\mu}_{\xi}-2\xi^{\nu}\theta^{\mu}\left[\delta\right]\right)\text{d}\mathcal{C}_{\mu\nu}-\int_{\mathcal{C}}\frac{1}{8\pi}\delta\Lambda\xi^{\mu}\text{d}\mathcal{C}_{\mu},
\end{equation}
where $\text{d}\mathcal{C}_{\mu\nu}=\left(\sqrt{-\mathfrak{g}}/\omega\right)^{-2/n}n_{[\mu}m_{\nu]}\omega\text{d}^{n-2}x$ is the Weyl invariant area element on the boundary $\partial\mathcal{C}$ (with $m_{\mu}$ being the unit normal to $\partial\mathcal{C}$ with respect to its embedding in $\mathcal{C}$ and $\text{d}^{n-2}x$ being the coordinate area element). Finally, if the Hamiltonian $H_{\xi}$ corresponding to the evolution along $\xi^{\mu}$ exists, then its perturbation $\delta H_{\xi}$ is given by equation~\eqref{dH1}. The main difference compared to the Hamiltonian perturbation in local, Diff-invariant theories is the presence of a volume integral proportional to $\delta\Lambda$~\cite{Wald:1993}. It can be traced back to $\Lambda$ being an integration constant rather than a fixed parameter in the Lagrangian. As the volume integral in general gives an infinite contribution to $\delta H_{\xi}$, the situations with nonzero $\Lambda$ or $\delta\Lambda$ need to be treated separately. Several such examples are discussed in the previous paper of the authors in the case of Weyl transverse gravity~\cite{Alonso:2022}. The results in general local, WTDiff-invariant gravitational theories are qualitatively the same and we briefly consider it in subsection~\ref{lambda}.

If $\Lambda=0$, i.e., the Hamiltonian $H_{\xi}$ does not contain a volume integral, we have the necessary and sufficient condition for the existence of  $H_{\xi}$ familiar from Diff-invariant theories of gravity~\cite{Wald:2000}
\begin{equation}
\label{condition}
\int_{\partial\mathcal{C}}\Omega^{\mu}\left[\pounds_{\xi},\delta\right]\xi^{\nu}\text{d}\mathcal{C}_{\mu\nu}=0.
\end{equation}
To see this, choose some solution of the equations of motion with $\Lambda=0$ and a vector field $\xi^{\mu}$ such that $H_{\xi}$ exists. Then consider two independent variations, $\delta_1g_{\mu\nu}$, $\delta_1\psi$, and $\delta_2g_{\mu\nu}$, $\delta_2\psi$, which solve the equations of motion with $\delta_1\Lambda=\delta_2\Lambda=0$. It holds $\left(\delta_1\delta_2-\delta_2\delta_1\right)H_{\xi}=0$. Expressing the variation of the Hamiltonian from equation~\eqref{dH1} and using the commutation of $\delta_1$ and $\delta_2$ on the antisymmetric tensor $Q^{\nu\mu}_{\xi}$ as well as the definition of $\Omega^{\mu}\left[\pounds_{\xi},\delta\right]$, we obtain the condition~\eqref{condition}.

\section{The first law of black hole mechanics and Wald entropy}
\label{physics}

An important application of the Noether charge formalism for gravitational theories is deriving the first law of black hole mechanics. Here, we first provide such a derivation for vacuum, asymptotically flat and stationary black hole solutions in WTDiff-invariant gravity. We then discuss the form of the first law in the presence of minimally coupled matter fields. Lastly, we invoke the definition of the Hawking temperature to identify the Wald entropy corresponding to a black hole horizon in local, WTDiff-invariant gravitational theories. We first consider the case $\Lambda=\delta\Lambda=0$. The cosmological constant contribution to the first law will be discussed in subsection~\ref{lambda} on the example of asymptotically anti-de Sitter spherically symmetric black hole spacetimes.

Before specialising to stationary black hole spacetimes, we first consider a more general case of an asymptotically flat spacetime possessing a Cauchy surface, $\mathcal{C}$, and a Killing vector, $\xi^{\mu}$. We consider a WTDiff-invariant definition of a Killing vector, i.e., \mbox{$\tilde{g}_{\rho(\nu}\tilde{\nabla}_{\mu)}\xi^{\rho}=0$}, implying $\pounds_{\xi}\tilde{g}_{\mu\nu}=0$ and $\pounds_{\xi}g_{\mu\nu}=g_{\mu\nu}\xi^{\rho}\partial_{\rho}\ln\left(\sqrt{-\mathfrak{g}}/\omega\right)$\footnote{Let us stress that while Killing vectors do not affect $\tilde{g}_{\mu\nu}$, this statement depends on the specific form of the metric. Therefore, if we evaluate Noether currents and charges derived without specifying the metric for a Killing vector, we generically obtain a nonvanishing result.}. Hence, the transformation generated by a Killing vector leaves the auxiliary metric invariant and acts as a pure Weyl transformation on the dynamical metric. This definition ensures that two spacetimes related by a Weyl transformation (which are physically identical in WTDiff-invariant gravity) have the same Killing vectors. In the unimodular gauge, our definition reduces to the one standard for Diff-invariant theories. Likewise, we consider a causal structure defined with respect to the auxiliary metric, $\tilde{g}_{\mu\nu}$, to ensure its invariance under the Weyl transformations.

In this asymptotically flat spacetime, we study an arbitrary perturbation of the metric which satisfies the equations of motion and does not spoil the asymptotic flatness. The symplectic current~\eqref{omega} corresponding to an arbitrary perturbation and a transformation generated by any Killing vector $\xi^{\mu}$ vanishes as $\pounds_{\xi}\tilde{g}_{\mu\nu}=0$. Hence, the perturbation of the Hamiltonian $\delta H_{\xi}$~\eqref{dH1} corresponding to the evolution along $\xi^{\mu}$ also vanishes. Equation~\eqref{dH1} then implies
\begin{equation}
\label{dH2}
\delta H_{\xi}=\int_{\partial\mathcal{C}}\left(\delta Q^{\nu\mu}_{\xi}-2\xi^{\nu}\theta^{\mu}\left[\delta\right]\right)\text{d}\mathcal{C}_{\mu\nu}=0,
\end{equation}
where the boundary $\partial\mathcal{C}$ of the Cauchy surface $\mathcal{C}$ in general consists of several components. One of them is an intersection $\partial\mathcal{C}_{\infty}$ of the Cauchy surface $\mathcal{C}$ with the spatial infinity, the others are internal components we collectively denote by $\partial\mathcal{C}_{\text{I}}$. These correspond, e.g. to intersections of $\mathcal{C}$ with black hole horizons. Equation~\eqref{dH2} now becomes
\begin{equation}
\label{dH killing}
\int_{\partial\mathcal{C}_{\infty}}\left(\delta Q^{\nu\mu}_{\xi}-2\xi^{\nu}\theta^{\mu}\left[\delta\right]\right)\text{d}\mathcal{C}_{\mu\nu}-\int_{\partial\mathcal{C}_{\text{I}}}\left(\delta Q^{\nu\mu}_{\xi}-2\xi^{\nu}\theta^{\mu}\left[\delta\right]\right)\text{d}\mathcal{C}_{\mu\nu}=0.
\end{equation}
The second term cannot be physically interpreted without specifying $\partial\mathcal{C}_{\text{I}}$. The first term corresponds to perturbations of the quantities measured in the asymptotic infinity. In particular, suppose that there is a timelike Killing vector, $t^{\mu}$, normalised so that \mbox{$\tilde{g}_{\mu\nu}t^{\mu}t^{\nu}=-1$} in the asymptotic infinity. Then, we can identify the contribution to $\delta H_{t}$ coming from $\partial\mathcal{C}_{\infty}$ with the perturbation of the total canonical energy of the spacetime~\cite{Wald:1993,Wald:1994}
\begin{equation}
\label{energy}
\delta E=\int_{\partial\mathcal{C}_{\infty}}\left(\delta Q^{\mu\nu}_{t}-2t^{\nu}\theta^{\mu}\left[\delta\right]\right)\text{d}\mathcal{C}_{\mu\nu}.
\end{equation}
Likewise, for a rotational  Killing vector $\varphi^{\mu}$, we obtain the perturbation of the total canonical angular momentum
\begin{equation}
\label{angular}
\delta J=-\int_{\partial\mathcal{C}_{\infty}}\delta Q^{\mu\nu}_{\varphi}\text{d}\mathcal{C}_{\mu\nu},
\end{equation}
where the overall minus sign gives us positive $J$. As $\varphi^{\mu}$ is orthogonal to $\text{d}\mathcal{C}_{\mu\nu}$, $\delta J$ lacks a contribution proportional to $\varphi^{\nu}\theta^{\mu}$. It is easy to check that the perturbations of the total canonical energy and angular momentum are Weyl invariant. In the unimodular gauge and for perturbations that do not change the metric determinant, $\delta\mathfrak{g}=0$, we recover the corresponding expressions valid for local, Diff-invariant theories of gravity~\cite{Wald:1993,Wald:1994}.

\subsection{WTDiff-invariant first law of black hole mechanics in vacuum}
\label{black holes}

We now restrict our attention to stationary, axisymmetric, vacuum spacetimes with a single black hole. By definition, such a spacetime possesses a time translational Killing vector $t^{\mu}$ and $n-3$ rotational Killing vectors $\varphi^{\mu}_{(i)}$. A combination of these vectors, $\xi^{\mu}=t^{\mu}+\Omega_{\mathcal{H}}^{(i)}\varphi^{\mu}_{(i)}$, with $\Omega_{\mathcal{H}}^{(i)}$ being constant angular velocities of the event horizon in directions $\varphi^{\mu}_{(i)}$, is a Killing vector orthogonal to the event horizon.  The inner boundary $\partial\mathcal{C}_{\text{I}}$ of the Cauchy surface $\mathcal{C}$ now consist of an intersection of $\mathcal{C}$ with the event horizon, $\mathcal{H}$, which we denote by $\partial\mathcal{C}_{\mathcal{H}}$. Equation~\eqref{dH killing} applied to this case yields
\begin{equation}
\delta E-\sum_{i=1}^{n-3}\Omega^{(i)}_{\mathcal{H}}\delta J_{(i)}-\int_{\partial\mathcal{C}_{\mathcal{H}}}\left[\delta Q^{\nu\mu}_{\xi}-2\xi^{\nu}\theta^{\mu}\right]\text{d}\mathcal{C}_{\mu\nu}=0,
\end{equation}
where we already identified the perturbations of the canonical energy~\eqref{energy} and the canonical angular momenta~\eqref{angular}. Since the Killing vector $\xi^{\mu}$ is orthogonal to the horizon, the second term in the integral over $\partial\mathcal{C}_{\mathcal{H}}$ vanishes. Likewise, terms proportional to $\xi^{\mu}$ in the Noether charge $Q^{\nu\mu}_{\xi}$~\eqref{charge} do not contribute to the integral (see~\cite{Jacobson:1994,Wald:1994} for details). In total, we have
\begin{equation}
\label{fl int}
\delta E-\sum_{i=1}^{n-3}\Omega^{(i)}_{\mathcal{H}}\delta J_{(i)}-2\int_{\partial\mathcal{C}_{\mathcal{H}}}\delta\left(E^{\nu\mu\rho}_{\quad\;\:\sigma}\tilde{\nabla}_{\rho}\xi^{\sigma}\right)\text{d}\mathcal{C}_{\mu\nu}=0.
\end{equation}
To further simplify the last term, we introduce a WTDiff-invariant definition of the surface gravity of the horizon
\begin{equation}
\label{kappa}
\kappa=\sqrt{g_{\mu\nu}g^{\rho\sigma}\tilde{\nabla}_{\rho}\xi^{\mu}\tilde{\nabla}_{\sigma}\xi^{\nu}}\Big\vert_{\partial\mathcal{C}_{\mathcal{H}}},
\end{equation}
and assume the validity of the zeroth law of black hole mechanics, i.e., that $\kappa$ is constant on the horizon. For proofs of this assumption in certain modified theories of gravity see, e.g.~\cite{Jacobson:1995,Ghosh:2020,Sang:2021,Bhattacharyya:2022} (these proofs directly carry over from Diff to WTDiff-invariant theories of gravity as the classical solutions are equivalent). Then, we can replace  $\tilde{\nabla}^{\rho}\xi^{\sigma}$ in equation~\eqref{fl int} by $\kappa\epsilon^{\rho\sigma}$, with $\epsilon^{\rho\sigma}$ being the bi-normal to the horizon~\cite{Jacobson:1994}. Finally, we arrive at the first law of black hole mechanics for vacuum, stationary, asymptotically flat spacetimes
\begin{equation}
\label{first law}
\delta E-\sum_{i=1}^{n-3}\Omega^{(i)}_{\mathcal{H}}\delta J_{(i)}-2\kappa\int_{\partial\mathcal{C}_{\mathcal{H}}}\delta\left(E^{\nu\mu\rho\sigma}\epsilon_{\rho\sigma}\right)\text{d}\mathcal{C}_{\mu\nu}=0,
\end{equation}
valid in any local, WTDiff-invariant theory of gravity. We recall that $E^{\nu\mu\rho\sigma}$ is given by equation~\eqref{entropy density}. As expected, in the unimodular gauge and for the metric determinant preserving perturbations, $\delta\mathfrak{g}=0$, equation~\eqref{first law} reduces to the first law for Diff-invariant theories~\cite{Wald:1994}.

\subsection{Wald entropy for WTDiff-invariant gravity}

We derived the first law of black hole mechanics in the setting of purely classical gravity. If we wish to relate it with the first law of thermodynamics, as is usually done in the literature~\cite{Wald:1993,Wald:1994,Iyer:1996}, we need to consider insights from the quantum field theory in a curved background. A well-known result of the quantum effects in black hole spacetime is emission of black body (Hawking) radiation by the black hole~\cite{Hawking:1975}. Since Hawking radiation comes from fluctuations of the matter fields, which are by construction Weyl invariant (and no quantum anomalies are associated with local Weyl transformations~\cite{Carballo:2015,Alvarez:2013b}), we expect its temperature to be Weyl invariant as well. Moreover, Hawking radiation is a kinematic effect independent of the gravitational dynamics~\cite{Visser:2003}. Therefore, the standard expression for the Hawking temperature, $T_{\text{H}}=\kappa/2\pi$, holds even in local, WTDiff-invariant theories of gravity, although  $\kappa$ must be understood as the Weyl invariant surface gravity~\eqref{kappa}.

Upon introducing the Hawking temperature, the last term in the first law~\eqref{first law} can be interpreted as $T_{\text{H}}\delta S$, with $S$ being Wald entropy of the horizon. For a black hole horizon in a stationary, asymptotically flat spacetime with a constant $\kappa$, its Wald entropy reads
\begin{equation}
\label{entropy}
S=-4\pi\int_{\partial\mathcal{C}_{\mathcal{H}}}E^{\mu\nu\rho\sigma}\epsilon_{\rho\sigma}\text{d}\mathcal{C}_{\mu\nu},
\end{equation}
where $E^{\mu\nu\rho\sigma}$ is given by equation~\eqref{entropy density}. In the unimodular gauge, we recover the prescription for Wald entropy valid in Diff-invariant gravitational theories~\cite{Wald:1993,Wald:1994}.

The first law of black hole thermodynamics in any local, WTDiff-invariant theory of gravity then has the usual form
\begin{equation}
\delta E-\sum_{i=1}^{n-3}\Omega^{(i)}_{\mathcal{H}}\delta J_{(i)}-T_{\text{H}}\delta S=0.
\end{equation}
However, it must be noted that the total energy, angular momenta, Hawking temperature, and Wald entropy are all Weyl invariant.

To conclude, we stress that the Hawking temperature does not naturally emerge in the Noether charge formalism. Rather, it must be introduced somewhat ad hoc by using insights from the quantum field theory in a curved background in an otherwise classical setting.
In this subsection, we do not aim to discuss the validity of such approach. Our only goal has been to obtain Wald entropy for WTDiff-invariant gravity and compare it with the well-known Diff-invariant result.

\subsection{Contributions of the cosmological constant and matter fields to the first law}
\label{lambda}

The most notable differences between WTDiff and Diff-invariant theories of gravity lie in the origin of the cosmological constant and in the status of local energy (non)conservation. Hence, in this section we discuss contributions to the first law of black hole mechanics coming from the cosmological constant and matter fields. Both cases are discussed in more detail for the special case of Weyl transverse gravity in~\cite{Alonso:2022}.

\paragraph{Cosmological constant contribution.}

We illustrate the contribution of a nonzero cosmological constant to the first law on the example of a static, spherically symmetric, asymptotically anti-de Sitter black hole spacetime. Any such spacetime possesses a time translational Killing vector field, $t^{\mu}$, and the black hole's horizon is a Killing horizon with respect to $t^{\mu}$. For a small vacuum perturbation of this spacetime that solves the equations of motion, equation~\eqref{dH1} implies
\begin{equation}
\label{dH lambda}
\delta H_{t}=\int_{\partial\mathcal{C}_{\infty}}\left(\delta Q^{\nu\mu}_{t}-2t^{\nu}\theta^{\mu}\left[\delta\right]\right)\text{d}\mathcal{C}_{\mu\nu}-\int_{\partial\mathcal{C}_{\mathcal{H}}}\delta Q^{\nu\mu}_{t}\text{d}\mathcal{C}_{\mu\nu}-\frac{1}{8\pi}\int_{\mathcal{C}}\delta\Lambda t^{\mu}\text{d}\mathcal{C}_{\mu}=0.
\end{equation}
We used that $t^{\mu}$ is orthogonal to the horizon.

The perturbation of the cosmological constant, $\delta\Lambda$, appears as an arbitrary integration constant in the process of solving the equations of motion for the metric perturbation. Hence, $\delta\Lambda$ is generically nonzero, i.e., the value of the cosmological constant differs between two solutions of the equations of motion related by a small perturbation\footnote{This is of course true even if $\Lambda=0$ in the original spacetime, we just set $\delta\Lambda=0$ in the previous subsections to simplify the discussion.}. Since the spatial volume of $\mathcal{C}$ is infinite, nonzero $\delta\Lambda$ leads to a divergent contribution to the first law. However, it is easy to see that the divergences are the same as in the first law in a pure anti-de Sitter spacetime whose cosmological constant is perturbed from $\Lambda$ to $\Lambda+\delta\Lambda$. Therefore, we can take anti-de Sitter spacetime as our reference background and require that the Noether current, charge and the symplectic potential vanish there. In other words, we define the physical quantities by subtracting their value in the purely anti-de Sitter spacetime (see subsection 4.3 of~\cite{Alonso:2022} for a more detailed discussion):
\begin{align}
\theta^{\mu}_{\text{phys}}\left[\delta\right]=&\theta^{\mu}\left[\delta\right]-\theta^{\mu}_{\text{AdS}}\left[\delta\right], \\
j^{\mu}_{t,\text{phys}}=&j^{\mu}_{t}-j^{\mu}_{t,\text{AdS}}=\tilde{\nabla}_{\nu}Q^{\nu\mu}_{t,\text{phys}}, \\
Q^{\nu\mu}_{t,\text{phys}}=&Q^{\nu\mu}_{t}-Q^{\nu\mu}_{t,\text{AdS}}.
\end{align}
The physical perturbation of the Hamiltonian then equals the difference of equation~\eqref{dH lambda} evaluated in the black hole spacetime and in the reference anti-de Sitter background
\begin{equation}
\label{dH phys}
\delta H_{t,\text{phys}}=\delta H_{t}-\delta H_{t,\text{AdS}}=\int_{\partial\mathcal{C}_{\infty}}\left(\delta Q^{\nu\mu}_{t,\text{phys}}-2t^{\nu}\theta^{\mu}_{\text{phys}}\left[\delta\right]\right)\text{d}\mathcal{C}_{\mu\nu}-\int_{\partial\mathcal{C}_{\mathcal{H}}}\delta Q^{\nu\mu}_{t,\text{phys}}\text{d}\mathcal{C}_{\mu\nu}.
\end{equation}
The integral over $\partial\mathcal{C}_{\infty}$ yields the perturbation of the total canonical energy, $\delta E$~\eqref{energy} (it is easy to realise that it contains no terms dependent on $\Lambda$ or $\delta\Lambda$ after the subtraction). Since the integral over $\partial\mathcal{C}_{\mathcal{H}}$ is finite even without any subtraction (the area of $\partial\mathcal{C}_{\mathcal{H}}$ is finite), we may evaluate both parts of $Q^{\nu\mu}_{t,\text{phys}}$ separately. The contribution of $Q^{\nu\mu}_{t}$ equals $2\kappa\int_{\partial\mathcal{C}_{\mathcal{H}}}\delta\left(E^{\nu\mu\rho\sigma}\epsilon_{\rho\sigma}\right)\text{d}\mathcal{C}_{\mu\nu}$ by the same argument as in the vacuum case. Finally, the integral of $Q^{\nu\mu}_{t,\text{AdS}}$ over $\partial\mathcal{C}_{\mathcal{H}}$ gives $\mathcal{V}_{n-1}r_{\mathcal{H}}^{n-1}\delta\Lambda/8\pi$, with $\mathcal{V}_{n-1}$ being the volume of $n-1$-dimensional flat space ball and $r_{\mathcal{H}}$ the black hole horizon's radius. In total, we have the following first law of mechanics for a static, spherically symmetric, asymptotically anti-de Sitter black hole spacetime
\begin{align}
\nonumber \delta E-2\kappa\int_{\partial\mathcal{C}_{\mathcal{H}}}\delta\left(E^{\nu\mu\rho\sigma}\epsilon_{\rho\sigma}\right)\text{d}\mathcal{C}_{\mu\nu}+\frac{1}{8\pi}\mathcal{V}_{n-1}r_{\mathcal{H}}^{n-1}\delta\Lambda=0.
\end{align}
As expected the first law is Weyl invariant. Its most interesting feature is the appearance of a term corresponding to the perturbation of the cosmological constant. The effects of this term has been extensively studied in the context of general relativity and it has been shown that the cosmological constant then plays a role analogous to that of pressure in standard thermodynamics (see, e.g.~\cite{Kastor:2009,Kubiznak:2015}). However, in general relativity (or any Diff-invariant theory) a varying cosmological constant must be introduced somewhat ad hoc, e.g. by appealing to its interpretation in terms of vacuum energy. In contrast, WTDiff-invariant gravity allows for cosmological constant variations already on the fully classical level (and regardless of its interpretation), making the study of contributions to the first law proportional to $\delta\Lambda$ more natural.

\paragraph{Matter field contributions.}

Another feature distinguishing WTDiff-invariant gravitational theories from the Diff-invariant ones is a possibility to break the local energy-momentum conservation (see equation~\eqref{non-conservation}). We show its impact on the first law on the example of a black hole spacetime filled with minimally coupled matter. The contribution of minimally coupled matter fields to the first law has been extensively discussed by the authors in the context of Weyl transverse gravity~\cite{Alonso:2022}. Going from Weyl transverse gravity to general local, WTDiff-invariant theories affects only the gravitational contributions to the first law, while the matter ones remain the same. We consider the case of a stationary, axisymmetric, asymptotically flat black hole spacetime containing any matter minimally coupled to gravity. Since we already analysed the gravitational part of the first law in subsection~\ref{black holes}, we can directly state the final result
\begin{align}
\nonumber &\delta E-\sum_{i=1}^{n-3}\Omega^{(i)}_{\mathcal{H}}\delta J^{(i)}_{\mathcal{H}}+2\kappa\int_{\partial\mathcal{C}_{\mathcal{H}}}\delta\left(E^{\nu\mu\rho\sigma}\epsilon_{\rho\sigma}\right)\text{d}\mathcal{C}_{\mu\nu} \\
\nonumber &-\int_{\mathcal{C}}\delta\left[\left(\sqrt{-\mathfrak{g}}/\omega\right)^{2k/n}T_{\nu}^{\;\:\mu}\right]U^{\nu}\text{d}\mathcal{C}_{\mu}+\frac{1}{2}\int_{\mathcal{C}}\left(\sqrt{-\mathfrak{g}}/\omega\right)^{-2/n}\mathring{T}^{\alpha\beta}\delta g_{\alpha\beta}\xi^{\mu}\text{d}\mathcal{C}_{\mu} \\
&+\int_{\mathcal{C}}\sum_{i=1}^{n-3}\Omega^{(i)}\delta\tilde{J}_{(i)}^{\mu}\text{d}\mathcal{C}_{\mu}+\int_{\mathcal{C}}\delta\mathcal{J}\xi^{\mu}\text{d}\mathcal{C}_{\mu}=0. \label{dH matter 2}
\end{align}
The first line is the same as in the vacuum case, except for $\delta J^{(i)}_{\mathcal{H}}$ being perturbations of the horizon (not total) angular momenta (both notions of course coincide in vacuum). The second line contains terms proportional to the energy-momentum tensor and its perturbation, with $U^{\nu}=t^{\nu}+\sum_{i=1}^{n-3}\Omega^{(i)}\varphi_{(i)}^{\nu}$, and $\Omega^{(i)}$ being the angular velocity of matter along $\varphi_{(i)}^{\nu}$. The first integral in the last line corresponds to the perturbations of the matter angular momenta. Finally, the second term in the last line quantifies the contribution of the local nonconservation of energy-momentum to the first law (for the meaning of $\mathcal{J}$, see equation~\eqref{non-conservation} and the accompanying discussion). Since, in Diff-invariant gravitational theories the generalised Bianchi identities enforce $\mathcal{J}=0$, this contribution represents the main difference between the WTDiff and Diff-invariant case.

The special case of a black hole spacetime filled with a perfect fluid discussed in~\cite{Alonso:2022} again generalises in the same way, with only the gravitational part affected by replacing Weyl transverse gravity with arbitrary local, WTDiff-invariant theories.

\section{First law of causal diamonds}
\label{conformal}

So far we were concerned with Hamiltonians corresponding to generators of transverse diffeomorphisms, which are local symmetries of WTDiff-invariant theories. However, more general vector fields are also of interest. For example, the Noether charge formalism for Diff-invariant gravity has been employed to derive the first law of causal diamonds~\cite{Jacobson:2015,Bueno:2017,Svesko:2019,Jacobson:2019}. Hence, one would expect that the same should be possible for WTDiff-invariant theories. This cannot be done by a direct application of the results of section~\ref{Noether}, since causal diamonds do not possess a timelike Killing vector. Instead, they have an isometry generated by a timelike conformal Killing vector defined for a given metric by the conformal Killing equation
\begin{equation}
\label{cke}
\delta_{\zeta}g_{\mu\nu}=\pounds_{\zeta}g_{\mu\nu}=2\nabla_{(\mu}\zeta_{\nu)}=\frac{1}{n}\nabla_{\rho}\zeta^{\rho}g_{\mu\nu}.
\end{equation}
A conformal Killing vector does not satisfy the transversality condition as $\tilde{\nabla}_{\mu}\zeta^{\mu}\ne0$. While equation~\eqref{cke} corresponds to a pure Weyl transformation which belongs to the WTDiff group, this statement is metric dependent. In other words, there does not exist a vector satisfying the conformal Killing equation for a general metric (in the same way a general metric has no true Killing vectors). Then, to derive the first law of causal diamonds, we need an expression for a perturbation of a Hamiltonian corresponding to evolution along any vector field, even if it does not generate a transformation belonging to the WTDiff group. Here, we obtain such an expression following the direct approach we previously applied in the case of Weyl transverse gravity~\cite{Alonso:2022}, although we modify it to obtain results valid even off shell.

While it would be possible to work with arbitrary local, WTDiff-invariant theories of gravity as before, the results turn out to be very difficult to interpret as there appears a large number of terms containing derivatives of $\tilde{\nabla}_{\mu}\zeta^{\mu}$, not present in the transverse case. Instead, we concentrate on Lagrangians that depend only on the auxiliary Riemann tensor and not on its derivatives. We also restrict our attention to minimally coupled matter fields. We choose this class of theories mainly because it is the choice typically made in works concerned with mechanics of causal diamonds~\cite{Jacobson:2015,Bueno:2017,Svesko:2019,Jacobson:2019}, which we discuss as an example in subsection~\ref{5.2}.

\subsection{Hamiltonian}
\label{5.1}

We start by discussing the derivation of an expression for a perturbation of a Hamiltonian corresponding to the evolution along any vector field. We consider a local gravitational Lagrangian of the form $L\left(\tilde{g}_{\mu\nu},\tilde{R}^{\mu}_{\;\:\nu\rho\sigma}\right)$ and a minimally coupled matter field described by the action~\eqref{matter}\footnote{If we were to include more than one matter field, the matter Lagrangian would be simply a sum of several terms of the form specified in equation~\eqref{matter}. We work with a single matter field for notational simplicity.}. Since we cannot define Noether currents and charges for general vector fields (they do not belong in the WTDiff symmetry group), we instead proceed by directly evaluating the symplectic potential and current. Integrating the symplectic current over a suitable Cauchy surface then yields the symplectic form and, hence, the Hamiltonian perturbation. Details of the derivation are rather technical and we leave them for Appendix~\ref{gen hamiltonian}.

The final expression for the perturbation $\delta H_{\zeta}$ of the Hamiltonian corresponding to the evolution along an arbitrary vector field $\zeta^{\mu}$ (assuming that both the original and perturbed spacetime satisfy the equations of motion) is
\begin{align}
\nonumber \delta H_{\zeta}=&\int_{\partial\mathcal{C}}\left[2\delta E^{\nu\mu\rho}_{\quad\;\:\sigma}\tilde{\nabla}_{\rho}\zeta^{\sigma}-4\zeta^{\sigma}\delta\left(\tilde{\nabla}_{\rho}E^{\nu\mu\rho}_{\quad\;\:\sigma}\right)+\tilde{\nabla}_{\nu}\delta Q^{\nu\mu}_{\psi,\zeta}-2\zeta^{\nu}\theta^{\mu}\left[\delta\right]\right]\text{d}\mathcal{C}_{\mu\nu}-\int_{\mathcal{C}}\frac{1}{8\pi}\delta\Lambda\zeta^{\mu}\text{d}\mathcal{C}_{\mu} \\
&+\int_{\mathcal{C}}\left(\Pi^{\mu}\left[\zeta,\delta\right]-\zeta^{\mu}\tilde{\nabla}_{\nu}\theta^{\nu}\left[\delta\right]+\frac{4}{n}\delta E^{\mu\nu\rho}_{\quad\;\:\nu}\tilde{\nabla}_{\rho}\tilde{\nabla}_{\lambda}\zeta^{\lambda}-\frac{4}{n}\delta\left(\tilde{\nabla}_{\rho}E^{\mu\nu\rho}_{\quad\;\:\nu}\right)\tilde{\nabla}_{\lambda}\zeta^{\lambda}\right)\text{d}\mathcal{C}_{\mu}.	\label{dH gen}
\end{align}
where the tensor $E_{\mu}^{\;\:\nu\rho\sigma}$ reduces to $E_{\mu}^{\;\:\nu\rho\sigma}=\partial L/\partial\tilde{R}^{\mu}_{\;\:\nu\rho\sigma}$ in this case, $\theta^{\mu}\left[\delta\right]$ is the usual symplectic potential, and $\Pi^{\mu}\left[\zeta,\delta\right]$ corresponds to a lengthy expression~\eqref{Pi} given in Appendix~\ref{gen hamiltonian}. Furthermore, $Q^{\nu\mu}_{\psi,\zeta}$ is an antisymmetric tensor built from the matter variables $\psi$ and their at most first derivatives, the auxiliary metric $\tilde{g}^{\mu\nu}$ and the vector field $\zeta^{\mu}$ (it contains no derivatives of $\zeta^{\mu}$).

The volume integral on the second line of equation~\eqref{dH gen} vanishes if $\tilde{\nabla}_{\mu}\zeta^{\mu}=0$, i.e., $\zeta^{\mu}$ generates a transverse diffeomorphism. In that case, we of course recover equation~\eqref{dH1}. Then, terms $2 E^{\nu\mu\rho}_{\quad\;\:\sigma}\tilde{\nabla}_{\rho}\zeta^{\sigma}-4\zeta^{\sigma}\tilde{\nabla}_{\rho}E^{\nu\mu\rho}_{\quad\;\:\sigma}$ and $Q^{\nu\mu}_{\psi,\zeta}$ correspond to the gravitational and matter part of the antisymmetric Noether charge tensor $Q^{\nu\mu}_{\zeta}$, respectively. We discuss the interpretation of the additional volume integral on the example of a causal diamond in the following subsection.

\subsection{Causal diamonds}
\label{5.2}

We now apply the general equation~\eqref{dH gen} to the case of a causal diamond in a flat spacetime. The main interest in causal diamonds comes from exploring the connection between gravity and thermodynamics, as they are locally constructed objects that exhibit thermodynamic properties analogous to those of black holes~\cite{Jacobson:2015,Bueno:2017,Svesko:2019,Jacobson:2019,Alonso:2020a,Alonso:2021}.

To construct a geodesic causal diamond centred at some spacetime point $P$, select an arbitrary unit timelike vector, $n^{\mu}\left(P\right)$, and a length scale, $l$. In every direction orthogonal to $n^{\mu}\left(P\right)$ we send out of $P$ geodesics of affine parameter length $l$ forming a spacelike geodesic ball $\Sigma_0$. The region causally determined by $\Sigma_0$ is known as a geodesic causal diamond.

We will work with a causal diamond constructed in a flat spacetime. In that case, we have an exact conformal isometry of the diamond generated by a conformal Killing vector, $\zeta^{\mu}$~\cite{Jacobson:2015}. In a coordinate system in which time is defined by $n\left(P\right)=\partial/\partial t$ and spatial coordinates are spherical, the conformal Killing vector reads, up to an arbitrary normalisation constant, $C$,
\begin{equation}
\label{conformal killing}
\zeta=C\left(\left(l^2-t^2-r^2\right)\frac{\partial}{\partial t}-2tr\frac{\partial}{\partial r}\right).
\end{equation}
The null boundary of the causal diamond is a conformal Killing horizon with respect to $\zeta^{\mu}$ and its cross-section with $\Sigma_0$, $\partial\Sigma_0$, is a bifurcate surface on which $\zeta^{\mu}$ vanishes. The surface gravity $\kappa$~\eqref{kappa} corresponding to $\zeta^{\mu}$ at $\partial\Sigma_0$ is $\kappa=2lC$.

Our aim is to derive the first law of causal diamonds, a formula governing small on-shell perturbations analogous to the first law of black hole mechanics. To do so, we consider a simultaneous perturbation of the metric and the matter fields around the flat background. Assuming that the perturbations satisfy the equations of motion, we obtain the perturbation of the Hamiltonian corresponding to the evolution along $\zeta^{\mu}$ by directly applying equation~\eqref{dH gen}, taking  the geodesic ball $\Sigma_0$ as the Cauchy surface. The Hamiltonian perturbation significantly simplifies since $\zeta^{\mu}$ vanishes on $\partial\Sigma_0$. Moreover, on $\Sigma_0$ we have $\tilde{\nabla}_{\mu}\zeta^{\mu}=0$, $\tilde{\nabla}_{\nu}\tilde{\nabla}_{\mu}\zeta^{\mu}=-2nC\delta^{t}_{\nu}$, and all the higher derivatives of $\tilde{\nabla}_{\mu}\zeta^{\mu}$ vanish identically. One can easily show that $\tilde{\nabla}^{\rho}\zeta^{\sigma}=\kappa\epsilon^{\rho\sigma}$, where $\epsilon^{\rho\sigma}$ denotes the bi-normal to $\partial\Sigma_0$. Furthermore, for flat spacetime the terms proportional to the Riemann tensor in the definition~\eqref{Pi} of $\Pi^{\mu}\left[\zeta,\delta\right]$ vanish and $E^{\sigma\nu\rho\mu}=\left(\sqrt{-\mathfrak{g}}/\omega\right)^{2/n}\left(g^{\mu\nu}g^{\rho\sigma}-g^{\mu\sigma}g^{\nu\rho}\right)/32\pi$, so that
\begin{equation}
\Pi^{\mu}\left[\zeta,\delta\right]=\frac{1}{16\pi}\frac{\kappa}{l}\delta^{t}_{\sigma}\delta\tilde{g}^{\sigma\mu}.
\end{equation}
In total, the general equation~\eqref{dH gen} becomes in this case
\begin{align}
\delta H_{\zeta}=&\int_{\partial\Sigma_0}2\kappa\epsilon^{\lambda\sigma}\tilde{g}_{\lambda\rho}\delta E^{\nu\mu\rho}_{\quad\;\:\sigma}\text{d}\mathcal{C}_{\mu\nu}-\int_{\Sigma_0}\frac{1}{8\pi}\delta\Lambda\zeta^{\mu}\text{d}\mathcal{C}_{\mu}+\int_{\Sigma_0}\frac{\kappa}{l}\delta^{t}_{\sigma}\left(\frac{1}{16\pi}\delta\tilde{g}^{\sigma\mu}-4\delta E^{\mu\nu\rho}_{\quad\;\:\nu}\right)\text{d}\mathcal{C}_{\mu}.
\end{align}

The perturbation of the Hamiltonian consists of two parts, one coming from the gravitational field and the other from the matter fields, $\delta H_{\zeta}=\delta H_{\text{g},\zeta}+\delta H_{\psi,\zeta}$. Evaluating $\Omega^{\mu}_{\psi}\left[\delta_{\zeta},\delta\right]$ proceeds along the same lines as in section 5 of~\cite{Alonso:2022}. Since flat spacetime implies a vanishing background value of the energy-momentum tensor, we have
\begin{equation}
\delta H_{\psi,\zeta}=\int_{\Sigma_0}\left[\left(\sqrt{-\mathfrak{g}}/\omega\right)^{2k/n}\delta T_{\nu}^{\;\:\mu}-\delta\mathcal{J}\delta^{\mu}_{\nu}\right]\zeta^{\nu}\text{d}\mathcal{C}_{\mu}. 
\end{equation}
The purely gravitational part of $\delta H_{\zeta}$, $\delta H_{\text{g},\zeta}$ vanishes because $\delta_{\zeta}\tilde{g}_{\mu\nu}=0$ implies $\Omega^{\mu}_{\text{g}}\left[\delta_{\zeta},\delta\right]=0$. However, the matter contribution, $\delta H_{\psi,\zeta}$, is generically nonvanishing~\cite{Iyer:1996,Jacobson:2019,Alonso:2022}. In total, the WTDiff-invariant first law of causal diamonds reads
\begin{align}
\nonumber &\int_{\Sigma_0}\left[\left(\sqrt{-\mathfrak{g}}/\omega\right)^{2k/n}\delta T_{\nu}^{\;\:\mu}-\delta\mathcal{J}\delta^{\mu}_{\nu}\right]\zeta^{\nu}\text{d}\mathcal{C}_{\mu}=\int_{\partial\Sigma_0}2\kappa\epsilon^{\lambda\sigma}\tilde{g}_{\lambda\rho}\delta E^{\nu\mu\rho}_{\quad\;\:\sigma}\text{d}\mathcal{C}_{\mu\nu} \\
&-\int_{\Sigma_0}\frac{1}{8\pi}\delta\Lambda\zeta^{\mu}\text{d}\mathcal{C}_{\mu}+\int_{\Sigma_0}\frac{\kappa}{l}\delta^{t}_{\sigma}\left(\frac{1}{16\pi}\delta\tilde{g}^{\sigma\mu}-4\delta E^{\mu\nu\rho}_{\quad\;\:\nu}\right)\text{d}\mathcal{C}_{\mu}. \label{diamond law}
\end{align}
The first term on the right hand side can be interpreted in terms of variation of Wald entropy~\eqref{entropy}. However, this interpretation again requires insights from quantum field theory and we will not discuss it further in this work.

The second term is proportional to the perturbation of the cosmological constant. While its presence has been discussed in the case of Diff-invariant gravity, it had to be justified by appealing to an interpretation of $\Lambda$ as a matter field rather than a fixed Lagrangian parameter~\cite{Jacobson:2019}. In our case, it appears naturally without any further assumptions, given the status of $\Lambda$ as a solution-dependent integration constant in WTDiff-invariant theories.

The last term corresponds to the perturbation of the so called generalised volume $W$ of $\Sigma_0$~\cite{Bueno:2017,Svesko:2019} 
\begin{equation}
\delta W=-\frac{8\pi}{n-2}\int_{\Sigma_0}\delta^{t}_{\sigma}\left(\frac{1}{16\pi}\delta\tilde{g}^{\sigma\mu}-4\delta E^{\mu\nu\rho}_{\quad\;\:\nu}\right)\text{d}\mathcal{C}_{\mu}.
\end{equation}
Of course, our expression for $\delta W$ is Weyl invariant and only reduces to the one defined in the context of Diff-invariant gravity in the unimodular gauge. In the case of Weyl transverse gravity, $\delta W$ equals $\left(\sqrt{-\mathfrak{g}}/\omega\right)^{2\left(n-1\right)/n}\delta\mathcal{V}_{n-1}$, with $\mathcal{V}_{n-1}$ being the geometric volume of $\Sigma_0$.

We stress that the generalised volume perturbation enters the first law in a different way than in Diff-invariant theories. There, the Hamiltonian is expressed entirely as a surface integral and $\delta W$ comes from the volume integral of the gravitational part of the symplectic current. However, the WTDiff-invariant gravitational symplectic current vanishes for conformal Killing vectors since $\delta_{\zeta}\tilde{g}_{\mu\nu}=0$. Instead, $\delta W$ appears in the first law from the volume integral present in $\delta_{\zeta}H$~\eqref{dH gen}.

To sum up, the first law of causal diamonds~\eqref{diamond law} we derived for WTDiff-invariant gravity is physically equivalent to the one valid in Diff-invariant theories~\cite{Bueno:2017,Svesko:2019}. However, it involves only Weyl invariant quantities and the generalised volume term appears in a different manner. Furthermore, it generically contains contributions corresponding to the cosmological constant perturbation and the local energy-momentum nonconservation.

\section{Discussion}
\label{discussion}

In this paper, we present two principal results on the Noether formalism for WTDiff-invariant theories of gravity and its applications. First, we have obtained the expressions for Noether currents and charges corresponding to transverse diffeomorphisms in any local, WTDiff-invariant theory of gravity. As an aside, we have also showed that the Noether current corresponding to the local Weyl symmetry identically vanishes in any local, WTDiff-invariant theory in an arbitrary spacetime dimension, generalising the earlier proof valid in four dimensions~\cite{Oda:2022}. We have then expressed the perturbation of the Hamiltonian corresponding to the evolution along any transverse diffeomorphism generator in terms of the Noether charge for this generator. In contrast to the Diff-invariant theories, here the Hamiltonian perturbation generically contains a volume integral corresponding to the perturbation of the cosmological constant.

Second, we have employed the Noether charge formalism to derive the first law of black hole mechanics in an arbitrary local WTDiff-invariant gravitational theory. In particular, we have explicitly found the form of the first law for a stationary, axisymmetric, asymptotically flat black hole spacetime filled with matter and for a static, spherically symmetric, asymptotically anti-de Sitter black hole spacetime. The resulting first law is in most respects physically equivalent to that valid in Diff-invariant gravity, although it involves Weyl invariant quantities. This is of course expected, since to every Diff-invariant theory there exist a WTDiff-invariant one with equivalent classical solutions. However, the WTDiff-invariant first law includes a contribution coming from a possible local nonconservation of energy-momentum, which is forbidden by the Diff invariance. Furthermore, it naturally contains a cosmological constant perturbation even on the level of fully classical physics. While a varying $\Lambda$ is sometimes considered even in Diff-invariant gravity, it is incompatible with its interpretation as a fixed parameter in the Lagrangian. Instead, some specific microscopic origin of $\Lambda$ needs to be assumed, at least implicitly.

As an additional result, we have also derived an expression for a perturbation of a Hamiltonian corresponding to a general vector field, including those that do not generate a transformation belonging to the WTDiff symmetry group. In this case, one does not have a conserved Noether current for the vector field and the Hamiltonian perturbation must be found by a direct calculation of the symplectic form. We have restricted our analysis to Lagrangians without derivatives of the auxiliary Riemann tensor and to minimally coupled matter fields, as the results become overly cumbersome in more general situations. We have found that the perturbation of the Hamiltonian contains an additional volume integral corresponding to the perturbation of the generalised volume. As an example, we then used the formula for the Hamiltonian perturbation to derive the first law of causal diamonds. We have again obtained a result physically equivalent to the one previously found for Diff-invariant theories (with the aforementioned differences regarding the energy-momentum nonconservation and the cosmological constant).

Aside from the specific results we have obtained, the formalism presented in this work offers a practical tool for further exploration of WTDiff-invariant theories of gravity. Among other things, it can be used to construct the solution phase space for any such theory (see~\cite{Adami:2021} and references therein), providing a possible way to study the (in)equivalence of WTDiff and Diff-invariant gravity. Our formalism might also be extended to study possible effects of the underlying dynamics of the background volume $n$-form, $\boldsymbol{\omega}$, on the first law of black hole mechanics. Lastly, it can serve to study the connection of WTDiff-invariant gravity and thermodynamics, either in the setting of black holes or locally constructed objects such as causal diamonds. In particular, it appears that thermodynamic arguments favour unimodular/WTDiff-invariant theories of gravity over the Diff-invariant ones~\cite{Tiwari:2006,Padmanabhan:2010,Alonso:2021}. The WTDiff-invariant Noether charge formalism applied to causal diamonds will allow a more rigorous examination of these claims (along the lines of~\cite{Bueno:2017,Svesko:2019}).

\section*{Acknowledgments}

The authors want to acknowledge Gerardo Garc\'{i}a-Moreno for helpful discussions. AA-S is supported by the ERC Advanced Grant No. 740209. ML is supported by the Charles University Grant Agency project No. GAUK 297721. This work is also partially supported by the Spanish Government through Project. No. MICINN PID2020-118159GB-C44.

\appendix
 
\section*{Appendices}

\section{Derivation of the symplectic potential}
\label{technical}

Here, we present the detailed derivation of the expression for the symplectic potential presented in subsection~\ref{potential}. Starting from an arbitrary variation of the Lagrangian~\eqref{action} carried out independently with respect to $g_{\mu\nu}$, $\tilde{R}^{\mu}_{\;\:\nu\rho\sigma}$ and $\psi$,
\begin{equation}
\delta L=\frac{\partial L}{\partial g_{\mu\nu}}\delta g_{\mu\nu}+\sum_{i=0}^{p}\frac{\partial L}{\partial\tilde{\nabla}_{(\alpha_1}...\tilde{\nabla}_{\alpha_i)}\tilde{R}^{\mu}_{\;\:\nu\rho\sigma}}\delta\tilde{\nabla}_{(\alpha_1}...\tilde{\nabla}_{\alpha_i)}\tilde{R}^{\mu}_{\;\:\nu\rho\sigma}+\sum_{i=0}^{q}\frac{\partial L}{\partial\tilde{\nabla}_{(\alpha_1}...\tilde{\nabla}_{\alpha_i)}\psi}\delta\tilde{\nabla}_{(\alpha_1}...\tilde{\nabla}_{\alpha_i)}\psi,
\end{equation}
we can use the following strategy to rewrite the terms containing derivatives of the auxiliary Riemann tensor
\begin{align}
\nonumber &\frac{\partial L}{\partial\tilde{\nabla}_{(\alpha_1}...\tilde{\nabla}_{\alpha_i)}\tilde{R}^{\mu}_{\;\:\nu\rho\sigma}}\delta\tilde{\nabla}_{(\alpha_1}...\tilde{\nabla}_{\alpha_i)}\tilde{R}^{\mu}_{\;\:\nu\rho\sigma} \\
\nonumber &=\frac{\partial L}{\partial\tilde{\nabla}_{(\alpha_1}...\tilde{\nabla}_{\alpha_i)}\tilde{R}_{\mu\nu\rho\sigma}}\tilde{\nabla}_{\alpha_1}\delta\tilde{\nabla}_{\alpha_2}...\tilde{\nabla}_{\alpha_i}\tilde{R}_{\mu\nu\rho\sigma}+\text{``terms proportional to $\tilde{\nabla}\delta g_{\mu\nu}$ and $\delta g_{\mu\nu}$''} \\
\nonumber &=\tilde{\nabla}_{\alpha_1}\left(\frac{\partial L}{\partial\tilde{\nabla}_{(\alpha_1}...\tilde{\nabla}_{\alpha_i)}\tilde{R}^{\mu}_{\;\:\nu\rho\sigma}}\delta\tilde{\nabla}_{\alpha_2}...\tilde{\nabla}_{\alpha_i}\tilde{R}^{\mu}_{\;\:\nu\rho\sigma}\right)-\tilde{\nabla}_{\alpha_1}\left(\frac{\partial L}{\partial\tilde{\nabla}_{(\alpha_1}...\tilde{\nabla}_{\alpha_i)}\tilde{R}^{\mu}_{\;\:\nu\rho\sigma}}\right)\delta\tilde{\nabla}_{\alpha_2}...\tilde{\nabla}_{\alpha_i}\tilde{R}^{\mu}_{\;\:\nu\rho\sigma} \\
\nonumber &+\tilde{\nabla}_{\alpha_1}\text{``terms proportional to $\delta g_{\mu\nu}$''}+\text{``terms proportional to $\delta g_{\mu\nu}$''} \\
\nonumber &=\tilde{\nabla}_{\alpha_1}\left(\frac{\partial L}{\partial\tilde{\nabla}_{(\alpha_1}...\tilde{\nabla}_{\alpha_i)}\tilde{R}^{\mu}_{\;\:\nu\rho\sigma}}\delta\tilde{\nabla}_{\alpha_2}...\tilde{\nabla}_{\alpha_i}\tilde{R}^{\mu}_{\;\:\nu\rho\sigma}\right)-\tilde{\nabla}_{\alpha_1}\left(\frac{\partial L}{\partial\tilde{\nabla}_{(\alpha_1}...\tilde{\nabla}_{\alpha_i)}\tilde{R}^{\mu}_{\;\:\nu\rho\sigma}}\right)\delta\tilde{\nabla}_{\alpha_2}...\tilde{\nabla}_{\alpha_i}\tilde{R}^{\mu}_{\;\:\nu\rho\sigma} \\
&+\tilde{\nabla}_{\alpha_1}\text{``terms proportional to $\delta g_{\mu\nu}$''}+\text{``terms proportional to $\delta g_{\mu\nu}$''}. \label{derivative rewrite}
\end{align}
We have obtained a lower derivative term, some Weyl covariant divergences, and terms proportional to $\delta g_{\mu\nu}$ which contribute to equations of motion. Terms with derivatives of the matter fields can be modified in an analogous way. A repeated use of this procedure then yields
\begin{align}
\nonumber \delta L=&\hat{A}^{\mu\nu}\delta g_{\mu\nu}+E_{\mu}^{\;\:\nu\rho\sigma}\delta\tilde{R}^{\mu}_{\;\:\nu\rho\sigma}+A_{\psi}\delta\psi \\
\nonumber &+\tilde{\nabla}_{\alpha_1}\bigg[\left(K^{\alpha_1\mu\nu}+2\tilde{g}^{\alpha_1\sigma}\tilde{\nabla}_{\rho}E_{\sigma}^{\;\:\mu\nu\rho}\right)\delta\tilde{g}_{\mu\nu}+\sum_{i=2}^{p}M^{\alpha_1\alpha_2...\alpha_i\;\:\nu\rho\sigma}_{\quad\quad\quad\;\mu}\delta\tilde{\nabla}_{(\alpha_2}...\tilde{\nabla}_{\alpha_i)}\tilde{R}^{\mu}_{\;\:\nu\rho\sigma} \\
&+\sum_{i=2}^{q}N^{\alpha_1\alpha_2...\alpha_k}\delta\tilde{\nabla}_{(\alpha_2}...\tilde{\nabla}_{\alpha_i)}\psi\bigg], \label{dL int}
\end{align}
where tensors $K^{\alpha_1\mu\nu}$, $M^{\alpha_1\alpha_2...\alpha_i\;\:\nu\rho\sigma}_{\quad\quad\quad\;\mu}$, and $N^{\alpha_1\alpha_2...\alpha_k}$ appear in the process of modifying the derivatives of $\delta\tilde{R}^{\mu}_{\;\:\nu\rho\sigma}$ and $\delta\psi$, using the procedure outlined in equation~\eqref{derivative rewrite}. Their explicit form is not relevant.

If the auxiliary Riemann tensor were an independent field, $\hat{A}^{\mu\nu}=0$ would be the gravitational equations of motion. Of course, $\delta\tilde{R}^{\mu}_{\;\:\nu\rho\sigma}$ actually depends on the variation of the auxiliary metric,
\begin{equation}
\delta\tilde{g}_{\mu\nu}=\left(\frac{\sqrt{-g}}{\omega}\right)^{-\frac{2}{n}}\left(\delta g_{\mu\nu}-\frac{1}{n}g_{\mu\nu}\frac{\delta g}{g}\right),
\end{equation}
and we can express it in terms of $\delta\tilde{g}_{\mu\nu}$ and its derivatives
\begin{equation}
E_{\mu}^{\;\:\nu\rho\sigma}\tilde{R}^{\mu}_{\;\:\nu\rho\sigma}=2E_{\mu}^{\;\:\nu\rho\sigma}\tilde{g}^{\mu\lambda}\tilde{\nabla}_{\lambda}\tilde{\nabla}_{\sigma}\delta\tilde{g}_{\nu\rho}+E_{\mu}^{\;\:\nu\rho\sigma}\tilde{R}^{\mu}_{\;\:\nu\rho\lambda}\tilde{g}^{\lambda\tau}\delta\tilde{g}_{\sigma\tau},
\end{equation}
where we have taken advantage of the symmetries of $E_{\mu}^{\;\:\nu\rho\sigma}$ (given in equation~\eqref{entropy density}) to simplify the variations. Then, the variation of the Lagrangian takes the form cited in the main text~\eqref{dL}. The traceless equations of motion read
\begin{align}
\nonumber \mathring{A}^{\mu\nu}+8\pi\mathring{T}_{\mu\nu}=& 8\pi\left(\frac{\sqrt{-g}}{\omega}\right)^{-\frac{2}{n}}\bigg[\hat{A}^{(\mu\nu)}+2E_{\iota}^{\;\:\rho\sigma(\mu\vert}\tilde{R}^{\iota\quad\vert\nu)}_{\;\:\rho\sigma}-\frac{1}{n}E_{\iota}^{\;\:\rho\sigma\lambda}\tilde{R}^{\iota}_{\;\:\rho\sigma\lambda}g^{\mu\nu} \\ &+4\tilde{\nabla}_{\sigma}\tilde{\nabla}^{\rho}E_{\rho}^{\;\:(\mu\nu)\sigma}-\frac{4}{n}g_{\iota\lambda}\left(\tilde{\nabla}_{\sigma}\tilde{\nabla}^{\rho}E_{\rho}^{\;\:\iota\lambda\sigma}\right)g^{\mu\nu}\bigg]=0.
\end{align}

\section{Hamiltonians corresponding to general vector fields}
\label{gen hamiltonian}

In this appendix, we present a detailed derivation of the perturbation~\eqref{dH gen} of the Hamiltonian corresponding to an arbitrary vector field $\zeta^{\mu}$.

Consider a local gravitational Lagrangian of the form $L\left(\tilde{g}_{\mu\nu},\tilde{R}^{\mu}_{\;\:\nu\rho\sigma}\right)$ and a minimally coupled matter field described by the action~\eqref{matter}. The traceless gravitational equations of motion read
\begin{align}
\nonumber 0=\mathring{A}^{\mu\nu}+8\pi\mathring{T}^{\mu\nu}=&-\left(\frac{\sqrt{-\mathfrak{g}}}{\omega}\right)^{\frac{2}{n}}E^{\mu\lambda\rho\sigma}\tilde{R}^{\nu}_{\;\:\lambda\rho\sigma}+2\tilde{\nabla}_{\rho}\tilde{\nabla}_{\sigma}\left[\left(\frac{\sqrt{-\mathfrak{g}}}{\omega}\right)^{\frac{2}{n}}E^{\mu\rho\sigma\nu}\right] \\
&+\frac{1}{n}\left(E^{\lambda\rho\sigma\tau}\tilde{R}^{\lambda\rho\sigma\tau}-2\tilde{\nabla}_{\rho}\tilde{\nabla}_{\sigma}E_{\lambda}^{\;\:\rho\sigma\lambda}\right)\tilde{g}^{\mu\nu}+8\pi\mathring{T}^{\mu\nu}, \label{traceless}
\end{align}
where the tensor $E_{\mu}^{\;\:\nu\rho\sigma}$ reduces to
\begin{equation}
E_{\mu}^{\;\:\nu\rho\sigma}=\frac{\partial L}{\partial\tilde{R}^{\mu}_{\;\:\nu\rho\sigma}}.
\end{equation}
The corresponding divergence-free equations~\eqref{divergence-free} (obtained by adding a scalar function multiplied by the auxiliary metric to equations~\eqref{traceless}) read
\begin{align}
\nonumber 0=A^{\mu\nu}=&-\left(\frac{\sqrt{-\mathfrak{g}}}{\omega}\right)^{\frac{2}{n}}E^{\mu\lambda\rho\sigma}\tilde{R}^{\nu}_{\;\:\lambda\rho\sigma}+2\tilde{\nabla}_{\rho}\tilde{\nabla}_{\sigma}\left[\left(\frac{\sqrt{-\mathfrak{g}}}{\omega}\right)^{\frac{2}{n}}E^{\mu\rho\sigma\nu}\right]+\frac{1}{2}L\tilde{g}^{\mu\nu} \\
&-\Lambda\tilde{g}^{\mu\nu}+8\pi\left(\frac{\sqrt{-\mathfrak{g}}}{\omega}\right)^{\frac{2\left(k+1\right)}{n}}T^{\mu\nu}-\mathcal{J}\tilde{g}^{\mu\nu},
\end{align}
where $\mathcal{J}$ is defined by equation~\eqref{non-conservation}. As always in WTDiff-invariant gravity, $\Lambda$ appears as an arbitrary integration constant.

The symplectic potential is easily worked out from the general expression~\eqref{theta}. Since the Lagrangian does not depend on derivatives of the auxiliary Riemann tensor, the terms containing tensors $M^{\mu\alpha_2...\alpha_i\;\:\nu\rho\sigma}=0$ do not appear and we have \mbox{$K^{\mu\nu\rho}=-2\tilde{\nabla}_{\sigma}\left[\left(\sqrt{-\mathfrak{g}}/\omega\right)^{2/n}E^{\mu\nu\rho\sigma}\right]$}. For the matter part of the symplectic potential we have $\theta^{\mu}_{\psi}=\left[\partial L_{\psi}/\partial\left(\partial_{\mu}\psi\right)\right]\delta\psi$ (a WTDiff invariant matter Lagrangian minimally coupled to gravity can only depend on $\tilde{g}^{\mu\nu}$, $\psi$ and $\partial_{\mu}\psi=\tilde{\nabla}_{\mu}\psi$). In total, we obtain
\begin{equation} \theta^{\mu}\left[\delta\right]=2\left(\frac{\sqrt{-\mathfrak{g}}}{\omega}\right)^{\frac{2}{n}}E^{\sigma\nu\rho\mu}\tilde{\nabla}_{\sigma}\delta\tilde{g}_{\nu\rho}-2\tilde{\nabla}_{\sigma}\left[\left(\frac{\sqrt{-\mathfrak{g}}}{\omega}\right)^{\frac{2}{n}}E^{\mu\nu\rho\sigma}\right]\delta\tilde{g}_{\nu\rho}+\frac{\partial L_{\psi}}{\partial\left(\partial_{\mu}\psi\right)}\delta\psi.
\end{equation}
The symplectic current corresponding to a transformation generated by any vector field $\zeta^{\mu}$ and an arbitrary variation of the metric and the matter fields reads
\begin{equation}
\Omega^{\mu}\left[\delta_{\zeta},\delta\right]=\delta_{\zeta}\theta^{\mu}\left[\delta\right]-\delta\theta^{\mu}\left[\delta_{\zeta}\right].
\end{equation}
As we already mentioned in subsection~\ref{potential}, $\delta_{\zeta}$ does not in general correspond to a Lie derivative. The reason is that $\delta_{\zeta}\boldsymbol{\omega}=0$ as the background volume form is nondynamical. However, $\pounds_{\zeta}\boldsymbol{\omega}=\boldsymbol{\omega}\tilde{\nabla}_{\mu}\zeta^{\mu}\ne0$ which equals zero only if $\zeta^{\mu}$ generates a transverse diffeomorphism, i.e., $\tilde{\nabla}_{\mu}\zeta^{\mu}=0$. Straightforward calculations show that the differences between $\delta_{\zeta}$ and $\pounds_{\zeta}$ acting on the building blocks of $\theta^{\mu}\left[\delta\right]$ are
\begin{align}
\left(\delta_{\zeta}-\pounds_{\zeta}\right)\frac{\sqrt{-\mathfrak{g}}}{\omega}=&\frac{\sqrt{-\mathfrak{g}}}{\omega}\tilde{\nabla}_{\mu}\zeta^{\mu}, \\
\left(\delta_{\zeta}-\pounds_{\zeta}\right)\tilde{\Gamma}^{\mu}_{\;\:\nu\rho}=&-\frac{1}{n}\left(\delta^{\mu}_{\nu}\delta^{\alpha}_{\rho}+\delta^{\mu}_{\rho}\delta^{\alpha}_{\nu}-g_{\nu\rho}g^{\mu\alpha}\right)\tilde{\nabla}_{\alpha}\tilde{\nabla}_{\lambda}\zeta^{\lambda} \\
\left(\delta_{\zeta}-\pounds_{\zeta}\right)\tilde{R}^{\mu}_{\;\:\nu\rho\sigma}=&\frac{2}{n}\left(\delta^{\alpha}_{\nu}\delta^{\mu}_{[\rho}\delta^{\beta}_{\sigma]}-g^{\mu\alpha}g_{\nu[\rho}\delta^{\beta}_{\sigma]}\right)\tilde{\nabla}_{\alpha}\tilde{\nabla}_{\beta}\tilde{\nabla}_{\lambda}\zeta^{\lambda} \\
\left(\delta_{\zeta}-\pounds_{\zeta}\right)\tilde{\nabla}_{\tau}\tilde{R}^{\mu}_{\;\:\nu\rho\sigma}=&\frac{2}{n}\left(\delta^{\alpha}_{\nu}\delta^{\mu}_{[\rho}\delta^{\beta}_{\sigma]}-g^{\mu\alpha}g_{\nu[\rho}\delta^{\beta}_{\sigma]}\right)\tilde{\nabla}_{\tau}\tilde{\nabla}_{\alpha}\tilde{\nabla}_{\beta}\tilde{\nabla}_{\lambda}\zeta^{\lambda}.
\end{align}
After some straightforward manipulations, we obtain a somewhat lengthy expression for $\Pi^{\mu}\left[\zeta,\delta\right]=\left(\delta_{\zeta}-\pounds_{\zeta}\right)\theta^{\mu}\left[\delta\right]$,
\begin{align}
\nonumber \Pi^{\mu}\left[\zeta,\delta\right]=&-\frac{8}{n}\left(\frac{\sqrt{-\mathfrak{g}}}{\omega}\right)^{\frac{2}{n}}F^{\sigma\nu\rho\mu\;\:\beta\alpha\gamma}_{\quad\;\;\;\:\alpha}\tilde{\nabla}_{\sigma}\tilde{\nabla}_{\beta}\tilde{\nabla}_{\gamma}\tilde{\nabla}_{\lambda}\zeta^{\lambda}\delta\tilde{g}_{\nu\rho} \\
\nonumber &+\frac{8}{n}\left(\frac{\sqrt{-\mathfrak{g}}}{\omega}\right)^{\frac{2}{n}}F^{\sigma\nu\rho\mu\;\:\beta\alpha\gamma}_{\quad\;\;\;\:\alpha}\tilde{\nabla}_{\beta}\tilde{\nabla}_{\gamma}\tilde{\nabla}_{\lambda}\zeta^{\lambda}\tilde{\nabla}_{\sigma}\delta\tilde{g}_{\nu\rho} \\
\nonumber &-\frac{8}{n}\left(\frac{\sqrt{-\mathfrak{g}}}{\omega}\right)^{\frac{2}{n}}G^{\sigma\nu\rho\mu\;\:\beta\alpha\gamma\;\:\tau\vartheta\omega}_{\quad\;\;\;\alpha\quad\;\;\xi}\tilde{\nabla}_{\sigma}\tilde{R}^{\xi}_{\;\:\tau\vartheta\omega}\tilde{\nabla}_{\beta}\tilde{\nabla}_{\gamma}\tilde{\nabla}_{\lambda}\zeta^{\lambda}\delta\tilde{g}_{\nu\rho} \\
\nonumber &-\frac{2}{n}\left(\frac{\sqrt{-\mathfrak{g}}}{\omega}\right)^{\frac{2}{n}}E^{\sigma\nu\rho\mu}\tilde{\nabla}_{\sigma}\tilde{\nabla}_{\lambda}\zeta^{\lambda}\delta\tilde{g}_{\nu\rho}+\frac{4}{n}\left(\frac{\sqrt{-\mathfrak{g}}}{\omega}\right)^{\frac{2}{n}}F^{\sigma\nu\rho\mu\;\:\beta\gamma\eta}_{\quad\;\;\;\:\alpha}\tilde{R}^{\alpha}_{\;\:\tau\vartheta\omega}\tilde{\nabla}_{\varkappa}\tilde{\nabla}_{\lambda}\zeta^{\lambda}\delta\tilde{g}_{\nu\rho} \\
\nonumber &\left(\delta^{\varkappa}_{\beta}\delta_{\sigma}^{\tau}\delta^{\vartheta}_{\gamma}\delta^{\omega}_{\eta}+\delta^{\varkappa}_{\eta}\delta^{\tau}_{\beta}\delta^{\vartheta}_{\gamma}\delta^{\omega}_{\sigma}+\delta^{\varkappa}_{\sigma}\delta^{\tau}_{\beta}\delta^{\vartheta}_{\gamma}\delta^{\omega}_{\eta}-g_{\sigma\beta}g^{\varkappa\tau}\delta^{\vartheta}_{\gamma}\delta^{\omega}_{\eta}-g_{\sigma\eta}g^{\varkappa\omega}\delta^{\tau}_{\beta}\delta^{\vartheta}_{\gamma}\right) \\
&+\frac{2k}{n}\frac{\partial L_{\psi}}{\partial\left(\partial_{\mu}\psi\right)}\delta\psi\tilde{\nabla}_{\lambda}\zeta^{\lambda}, \label{Pi}
\end{align}
where
\begin{align}
F^{\;\:\nu\rho\sigma\;\:\beta\gamma\eta}_{\mu\quad\;\alpha}=&\frac{\partial E_{\mu}^{\;\:\nu\rho\sigma}}{\partial \tilde{R}^{\alpha}_{\;\:\beta\gamma\eta}}=\frac{\partial^2 L}{\partial\tilde{R}^{\alpha}_{\;\:\beta\gamma\eta}\partial\tilde{R}^{\mu}_{\;\:\nu\rho\sigma}}, \\
G^{\;\:\nu\rho\sigma\;\:\beta\gamma\eta\;\:\tau\vartheta\omega}_{\mu\quad\alpha\quad\;\;\xi}=&\frac{\partial^2 E_{\mu}^{\;\:\nu\rho\sigma}}{\partial\tilde{R}^{\xi}_{\;\:\tau\vartheta\omega}\partial\tilde{R}^{\alpha}_{\;\:\beta\gamma\eta}}=\frac{\partial^3 L}{\partial\tilde{R}^{\xi}_{\;\:\tau\vartheta\omega}\partial\tilde{R}^{\alpha}_{\;\:\beta\gamma\eta}\partial\tilde{R}^{\mu}_{\;\:\nu\rho\sigma}}.
\end{align}
We now use that $\delta_{\zeta}\theta^{\mu}\left[\delta\right]=\pounds_{\zeta}\theta^{\mu}\left[\delta\right]+\Pi^{\mu}\left[\zeta,\delta\right]$. Expanding $\pounds_{\zeta}\theta^{\mu}\left[\delta\right]$ then yields
\begin{align}
\nonumber \delta_{\zeta}\theta^{\mu}\left[\delta\right]=&\pounds_{\zeta}\theta^{\mu}\left[\delta\right]+\Pi^{\mu}\left[\zeta,\delta\right]=\zeta^{\nu}\tilde{\nabla}_{\nu}\theta^{\mu}\left[\delta\right]-\theta^{\nu}\left[\delta\right]\tilde{\nabla}_{\nu}\zeta^{\nu}+\Pi^{\mu}\left[\zeta,\delta\right] \\
\nonumber =&-2\tilde{\nabla}_{\nu}\left(\theta^{[\nu}\left[\delta\right]\zeta^{\mu]}\right)+\zeta^{\mu}\frac{1}{16\pi}\left(\frac{\sqrt{-\mathfrak{g}}}{\omega}\right)^{\frac{2}{n}}\left(\mathring{A}^{\alpha\beta}+8\pi\mathring{T}^{\alpha\beta}\right)\delta g_{\alpha\beta} \\
&+\zeta^{\mu}A_{\psi}\delta\psi-\zeta^{\mu}\delta L-\theta^{\mu}\left[\delta\right]\tilde{\nabla}_{\nu}\zeta^{\nu}+\Pi^{\mu}\left[\zeta,\delta\right],
\end{align}
where we used the general equation~\eqref{dL general} to express $\tilde{\nabla}_{\nu}\theta^{\nu}\left[\delta\right]$ in the second equality.

Regarding the second term in $\Omega^{\mu}\left[\delta_{\zeta},\delta\right]$, a direct calculation leads to
\begin{align}
\nonumber -\delta\theta^{\mu}\left[\delta_{\zeta}\right]=&-\frac{1}{8\pi}\left(\frac{\sqrt{-\mathfrak{g}}}{\omega}\right)^{\frac{2}{n}}A^{\;\:\mu}_{\nu}\zeta^{\nu}+\zeta^{\mu}\delta L-\delta\left(\zeta\cdot A_{\psi}\cdot\psi\right)^{\mu} \\
\nonumber &+\frac{4}{n}\delta E^{\mu\nu\rho}_{\quad\;\:\nu}\tilde{\nabla}_{\rho}\tilde{\nabla}_{\lambda}\zeta^{\lambda}-\frac{4}{n}\delta\left(\tilde{\nabla}_{\rho}E^{\mu\nu\rho}_{\quad\;\:\nu}\right)\tilde{\nabla}_{\lambda}\zeta^{\lambda}\\
&+\tilde{\nabla}_{\nu}\left[2\tilde{\nabla}_{\rho}\zeta^{\sigma}\delta E^{\nu\mu\rho}_{\quad\;\:\sigma}-4\zeta^{\sigma}\delta\left(\tilde{\nabla}_{\rho}E^{\nu\mu\rho}_{\quad\;\:\sigma}\right)\right]+\tilde{\nabla}_{\nu}\delta Q^{\nu\mu}_{\psi,\zeta},
\end{align}
where $Q^{\nu\mu}_{\psi,\zeta}$ is an antisymmetric tensor that arises from the matter part of $\theta^{\mu}\left[\delta_{\zeta}\right]$ which depends on $\zeta^{\mu}$ but not its derivatives\footnote{The calculations concerning the matter part of the symplectic current proceeds along the same lines as in the section 5 of~\cite{Alonso:2022}.}. If $\zeta^{\mu}$ generates a transverse diffeomorphism, the last line can be identified as the Weyl covariant divergence of the corresponding Noether charge (this is clear from the comparison with equation~\eqref{dj}).

Altogether, we have for the symplectic current $\Omega^{\mu}\left[\delta_{\zeta},\delta\right]$
\begin{align}
\nonumber \Omega^{\mu}\left[\delta_{\zeta},\delta\right]=&\tilde{\nabla}_{\nu}\left[2\tilde{\nabla}_{\rho}\zeta^{\sigma}\delta E^{\nu\mu\rho}_{\quad\;\:\sigma}-4\zeta^{\sigma}\delta\left(\tilde{\nabla}_{\rho}E^{\nu\mu\rho}_{\quad\;\:\sigma}\right)\right]-2\tilde{\nabla}_{\nu}\left(\theta^{[\nu}\left[\delta\right]\zeta^{\mu]}\right)+\tilde{\nabla}_{\nu}\delta Q^{\nu\mu}_{\psi,\zeta} \\
\nonumber & +\Pi^{\mu}\left[\zeta,\delta\right]-\theta^{\mu}\left[\delta\right]\tilde{\nabla}_{\nu}\zeta^{\nu}+\frac{4}{n}\delta E^{\mu\nu\rho}_{\quad\;\:\nu}\tilde{\nabla}_{\rho}\tilde{\nabla}_{\lambda}\zeta^{\lambda}-\frac{4}{n}\delta\left(\tilde{\nabla}_{\rho}E^{\mu\nu\rho}_{\quad\;\:\nu}\right)\tilde{\nabla}_{\lambda}\zeta^{\lambda}\zeta^{\mu} \\
\nonumber &+\frac{1}{16\pi}\left(\frac{\sqrt{-\mathfrak{g}}}{\omega}\right)^{\frac{2}{n}}\left(\mathring{A}^{\alpha\beta}+8\pi\mathring{T}^{\alpha\beta}\right)\delta g_{\alpha\beta}-\frac{1}{8\pi}\left(\frac{\sqrt{-\mathfrak{g}}}{\omega}\right)^{\frac{2}{n}}A^{\;\:\mu}_{\nu}\zeta^{\nu}\\
&+\zeta^{\mu}A_{\psi}\delta\psi-\delta\left(\zeta\cdot A_{\psi}\cdot\psi\right)^{\mu}.
\end{align}

So far, we have been working off shell. Assuming that both the original and the perturbed spacetime satisfy the equations of motion, it holds
\begin{align}
\nonumber &\Omega^{\mu}\left[\delta_{\zeta},\delta\right]=\tilde{\nabla}_{\nu}\left[2\tilde{\nabla}_{\rho}\zeta^{\sigma}\delta E^{\nu\mu\rho}_{\quad\;\:\sigma}-4\zeta^{\sigma}\delta\left(\tilde{\nabla}_{\rho}E^{\nu\mu\rho}_{\quad\;\:\sigma}\right)\right]+\tilde{\nabla}_{\nu}\delta Q^{\nu\mu}_{\psi,\zeta}-2\tilde{\nabla}_{\nu}\left(\theta^{[\nu}\left[\delta\right]\zeta^{\mu]}\right) \\
&-\frac{1}{8\pi}\zeta^{\mu}\delta\Lambda+\Pi^{\mu}\left[\zeta,\delta\right]-\theta^{\mu}\left[\delta\right]\tilde{\nabla}_{\nu}\zeta^{\nu}+\frac{4}{n}\delta E^{\mu\nu\rho}_{\quad\;\:\nu}\tilde{\nabla}_{\rho}\tilde{\nabla}_{\lambda}\zeta^{\lambda}-\frac{4}{n}\delta\left(\tilde{\nabla}_{\rho}E^{\mu\nu\rho}_{\quad\;\:\nu}\right)\tilde{\nabla}_{\lambda}\zeta^{\lambda}.
\end{align}
Finally, integrating over a suitable Cauchy surface $\mathcal{C}$ yields the symplectic form $\Omega\left[\delta_{\zeta},\delta\right]$ and, therefore, the perturbation of the Hamiltonian corresponding to the evolution along $\zeta^{\mu}$~\eqref{dH gen}.

\end{document}